\documentstyle[11pt,amsmath,amssymb,epsf,epic, color]{article}

\numberwithin{equation}{section}

\newcommand{\fa}{{\frak A}}
\newcommand{\fm}{{\frak M}}
\newcommand{\mm}{{\frak M}}
\newcommand{\un}{\bbbone}
\newcommand{\ra}{\rightarrow}
\newcommand{\tr}{{\rm Tr}}

\newcommand{\bra}{\langle}
\newcommand{\ket}{\rangle}
\newcommand{\fh}{{\mathfrak h}}
\newcommand{\sinc}{{\rm sinc}}
\renewcommand{\i}{{\rm i}}

\newcommand{\be}{\begin{equation}}
\newcommand{\ee}{\end{equation}}
\newcommand{\bea}{\begin{eqnarray}}
\newcommand{\eea}{\end{eqnarray}}

\newcommand{\beq}{\begin{equation}}
\newcommand{\eeq}{\end{equation}}
\newcommand{\bex}{\begin{example}}
\newcommand{\eex}{\end{example}}
\newcommand{\ber}{\begin{remark}}
\newcommand{\eer}{\end{remark}}
\def\bel{\begin{lemma}}
\def\eel{\end{lemma}}
\def\bet{\begin{theorem}}
\def\eet{\end{theorem}}
\def\bed{\begin{definition}}
\def\eed{\end{definition}}

\newcommand{\eps}{\epsilon}

\newcommand{\ce}{{\cal E}}

\newcommand{\cc}{{\cal C}}

\newcommand{\cs}{{\cal S}}

\newcommand{\e}{{\rm e}}

\renewcommand{\d}{{\rm d}}

\newcommand{\s}{{\cal S}}

\newcommand{\ffi}{\varphi}

\newcommand{\ep}{\hfill  {\vrule height 10pt width 8pt depth 0pt}}

\newcommand{\grintl}{[\kern-.18em [}
\newcommand{\grintr}{]\kern-.18em ]}
\newcommand{\ds}{\displaystyle}

\newcounter{resultcounter}[section]
\renewcommand{\theresultcounter}{\arabic{section}.\arabic{resultcounter}}

\newtheorem{thm}[resultcounter]{Theorem}
\newtheorem{lem}[resultcounter]{Lemma}
\newtheorem{prop}[resultcounter]{Proposition}
\newtheorem{cor}[resultcounter]{Corollary}
\newtheorem{definition}[resultcounter]{Definition}

\def\bed{\begin{definition}}
\def\eed{\end{definition}}

\def\one{{\mathchoice {\rm 1\mskip-4mu l} {\rm 1\mskip-4mu l} {\rm 1\mskip-4.5mu l} {\rm 1\mskip-5mu l}}}

 \def\cB{{\cal B}} \def\cC{{\cal C}}
\def\cD{{\cal D}} \def\cE{{\cal E}} 
 \def\cH{{\cal H}}

  \def\cR{{\cal R}}
\def\cS{{\cal S}}

\newcommand{\R}{{\mathbb R}}
\newcommand{\N}{{\mathbb N}}

\newcommand{\C}{{\mathbb C}}
\newcommand{\Z}{{\mathbb Z}}

\newcommand{\I}{{\mathbb I}}

\newcommand{\ri}{{\rm i}}
\newcommand{\fer}[1]{(\ref{#1})}
\newcommand{\scalprod}[2]{\left\langle {#1}, {#2}\right\rangle}
\newcommand{\bbbone}{\mathchoice {\rm 1\mskip-4mu l} {\rm 1\mskip-4mu l}
{\rm 1\mskip-4.5mu l} {\rm 1\mskip-5mu l}}

\newcommand{\rx}{{\mathbb R}}

\renewcommand{\I}{\bbbone}

\begin{document}
\title{Repeated and continuous interactions\\
 in open quantum systems}
\author{ Laurent Bruneau\footnote{
Laboratoire AGM,
Universit\'e de Cergy-Pontoise, Site Saint-Martin, BP 222,
 95302 Cergy-Pontoise, France. Email: laurent.bruneau@u-cergy.fr, http://www.u-cergy.fr/bruneau
}\ , Alain Joye\footnote{ Institut Fourier, UMR 5582,
CNRS-Universit\'e de Grenoble I BP 74, 38402 Saint-Martin
d'H\`eres, France. Email: Alain.Joye@ujf-grenoble.fr,
http://www-fourier.ujf-grenoble.fr/\ $\widetilde{}$\,joye }\
\footnote{Supported partially from Insitute for Mathematical
Sciences, National University of Singapore, through the program
``Mathematical Horizons for Quantum Physics'', during which parts
of this work have been performed. Partially supported by the
Minist\`ere Fran\c cais des Affaires \'Etrang\`eres through a
s\'ejour scientifique haut niveau.}\ , Marco
Merkli\footnote{Department of Mathematics, Memorial University of
Newfoundland, Canada. Supported by NSERC under Discovery Grant
205247. Email: merkli@mun.ca, http://www.math.mun.ca/\
$\widetilde{}$\,merkli/}\ \footnote{Supported partially from
Insitute for Mathematical Sciences, National University of
Singapore, through the program ``Mathematical Horizons for Quantum
Physics'', during which parts of this work have been performed.
Partially supported by the Minist\`ere Fran\c cais des Affaires
\'Etrang\`eres through a s\'ejour scientifique haut niveau.} }

\date{\today}

\maketitle
\vspace{-1cm}
\begin{abstract}
We consider a finite quantum system $\cs$ coupled to two
environments of different nature. One is a heat reservoir $\cR$
(continuous interaction) and the other one is a chain $\cc$ of
independent quantum systems $\ce$ (repeated interaction). The
interactions of $\cs$ with $\cR$ and $\cc$ lead to two
simultaneous dynamical processes. We show that for generic such
systems, any initial state approaches an asymptotic state in the
limit of large times. We express the latter in terms of the
resonance data of a reduced propagator of $\cs+\cR$ and show that
it satisfies a second law of thermodynamics. We analyze a model
where both $\cs$ and $\ce$ are two-level systems and obtain the
asymptotic state explicitly (lowest order in the interaction
strength). Even though $\cR$ and $\cc$ are not direcly coupled, we
show that they exchange energy, and we find the dependence of this
exchange in terms of the thermodynamic parameters.

We formulate the problem in the framework of $W^*$-dynamical
systems and base the analysis on a combination of spectral
deformation methods and repeated interaction model techniques. We
do not use master equation approximations.
\end{abstract}

\thispagestyle{empty}
\setcounter{page}{1}
\setcounter{section}{1}


\setcounter{section}{0}

\section{Introduction}

Over the last years, the rigorous study of equilibrium and
non-equilibrium quantum systems has received much and renewed
attention. While this topic of fundamental interest has a long
tradition in physics and mathematics, conventionally explored via
master equations \cite{BP, AJP}, dynamical semi-groups \cite{AL,
AJP} and algebraic scattering theory \cite{R, FMUe}, many recent
works focus on a quantum resonance theory approach. The latter has
been applied successfully to systems close to equilibrium
\cite{JP, msb1, msb2, msb3} and far from equilibrium \cite{jp2002,
MMS}. In both situations, one of the main questions is the (time-)
asymptotic behaviour of a quantum system consisting of a subsystem
$\s$ interacting with one or several other subsystems, given by
thermal reservoirs ${\cal R}_1,\ldots,{\cal R}_n$. It has been
shown that if ${\cal S}+{\cal R}$ starts in a state in which the
reservoir is in a thermal state at temperature $T>0$ far away from
the system $\cal S$, then ${\cal S}+{\cal R}$ converges to the
joint equilibrium state at temperature $T$, as time
$t\rightarrow\infty$. This phenomenon is called {\it return to
equilibrium}. (See also \cite{Mcond} for the situation where
several equilibrium states at a fixed temperature coexist.) In
case ${\cal S}$ is in contact with several reservoirs having
different temperatures (or different other macroscopic
properties), the whole system converges to a {\it non-equilibrium
stationary state} (NESS). The success of the resonance approach is
measured not only by the fact that the above-mentioned phenomena
can be described rigorously and quantitatively (convergence
rates), but also by that the asymptotic states can be constructed
(via perturbation theory in the interaction) and their physical
and mathematical structure can be examined explicitly (entropy
production, heat- and matter fluxes). One of the main advantages
of this method over the usual master equation approach (and the
related van Hove limit) is that it gives a perturbation theory of
the dynamics which is {\it uniform} in time $t\geq 0$. While the
initial motivation for the development of the dynamical resonance
theory was the investigation of the time-asymptotics, the method
is becoming increasingly refined. It has been extended to give a
precise picture of the dynamics of open quantum systems for all
times $t\geq 0$, with applications to the phenomena of
decoherence, disentanglement, and their relation to thermalization
\cite{msb2, msb1, msb3, GM}. An extension to systems with rather
arbitrary time-dependent Hamiltonians has been presented in
\cite{MStarr} (see also \cite{AF} for time-periodic systems). A
further direction of development is a quantum theory of linear
response and of fluctuations \cite{jpo}.

 In certain physical
setups, the  reservoir has a structure of
 {\it a chain of independent elements}, ${\cal C}={\cal E}_1
 +{\cal E}_2+\cdots$. An example of such a system is the so-called ``one-atom maser'' \cite{MWM}, where ${\cal S}$ describes the modes of the electromagnetic field in a cavity, interacting with a beam $\cal C$ of atoms ${\cal E}_j$, shot one by one into  the cavity and interacting for a duration $\tau_j>0$ with it. A mathematical treatment of the one-atom maser is provided in \cite{BPi}. Another instance of the use of such systems is the construction of reservoirs made of ``quantum noises'' by means of adequate scaling limits of the characteristics of the chain ${\cal C}$ and its coupling with ${\cal S}$, which lead to certain types of master equations \cite{A, AP, AJP, AJ1, AJ2}. The central feature of such systems is that $\cal S$ interacts successively with independent elements ${\cal E}_j$ constituting a reservoir. This independence implies a markovian property which simplifies the mathematical treatment considerably. In essence it enables one to express the dynamics of ${\cal S}$ at time $t=\tau_1+\cdots+\tau_N$ by a propagator of product form $M_1(\tau_1)\cdots M_N(\tau_N)$, where each $M_j(\tau_j)$ encodes the dynamics of $\cal S$ with a fixed element ${\cal E}_j$. In case each element ${\cal E}_j$ is physically the same and each interaction is governed by a fixed duration $\tau$ (and a fixed interaction operator), the dynamics is given by $M(\tau)^N$ and the asymptotics is encoded in the spectrum of the {\it reduced dynamics operator} $M(\tau)$ \cite{bjm1}. An analysis for non-constant interactions is more involved. It has been carried out in \cite{bjm2, bjm3}
 for systems with random characteristics (e.g. random
 interaction times). See also \cite{PN} for related issues.
 In both the deterministic and the random settings,
 the system approaches a limit state as
 $t\rightarrow\infty$, called a {\it repeated interaction
  asymptotic state} (RIAS), whose physical and mathematical
 properties have been investigated explicitly. \\

In the present work we make the synthesis of the above two
situations. We consider a system $\cal S$ interacting with two
environments of distinct nature. On the one hand, $\cal S$ is
coupled in the repeated interaction way to a chain ${\cal C}={\cal
E}+{\cal E}+\cdots$, and on the other hand, ${\cal S}$ is in
continuous contact with a heat reservoir ${\cal R}$. It is assumed
that $\cal C$ and $\cal R$ do not interact directly. Our goal is
to construct the asymptotic state of the system and to analyze its
physical properties. In particular, we present in Section
\ref{ssm1} our results on the convergence to, and form of the
asymptotic state, in Section \ref{sectthermo} the thermodynamic
properties of it, and in Section \ref{ssec:explexa} we present the
analysis of an explicit model.


\subsection{Description of the system}\label{sec:descripintro}

The following is a unified description of $\s$, $\cR$, $\cc$ in
the language of algebraic quantum statistical mechanics (we refer
the reader to e.g. \cite{P} for a more detailed exposition). The
Hilbert spaces of states of each of the subsystems $\#=\s, \cR,
\ce$ are  ${\cal H}_{\#}$. The respective observables form von
Neumann algebras $\mm_{\# }\subset {\cal B}(\cH_{\#})$. We assume
that $\dim\cH_\s<\infty$ and $\dim\cH_{\ce}$ may be finite or
infinite. $\cR$ being a reservoir, its Hilbert space is assumed to
be infinite-dimensional, $\dim\cH_\cR=\infty$. The free dynamics
of each constituent is generated by Liouville operators $L_{\#}$,
i.e., the Heisenberg evolution of an observable $A\in\mm_{\#}$ at
time $t$ is given by $\e^{\i t L_{\#}} A \e^{-\i t L_{\#}}$. In
each Hilbert space we pick a normalized {\it reference state}
$\Psi_\#$ which determines the macroscopic properties of the
systems.\footnote{In other words, it determines the folium of
normal states. If the Hilbert space is finite-dimensional then the
set of normal states is unique, but for infinite systems different
classes of normal states are determined by different macroscopic
parameters, such as the temperature.} These reference vectors are
invariant, $\e^{-\i t L_\#}\Psi_\#=\Psi_\#$, and they are cyclic
and separating for $\mm_\#$ \cite{BR}. Typically, the $\Psi_\#$
are chosen to be the equilibrium states at any fixed temperature
$T_\#>0$. The Hilbert space $\cH_\cc$ of the chain is the infinite
tensor product of factors $\cH_\ce$, taken with respect to the
stabilizing sequence $\Psi_\cc=\otimes_{j\geq 1}\Psi_\ce$.

In summary, the non-interacting system is given by a Hilbert space
\begin{equation*}
\cH = \cH_\cS\otimes\cH_\cR\otimes\cH_\cc
\label{-2}
\end{equation*}
and its dynamics is generated by the Liouvillian
\begin{equation}
L_0 = L_\cS+ L_\cR+\sum_{k\geq 1} L_{\ce_k}.
\label{-3}
\end{equation}
Here we understand that $L_{\cE_k}$ acts as the fixed operator $L_\ce$ on the $k$-th factor of $\cH_\cc$, and we do not display obvious factors $\bbbone$.

The operators governing the couplings between $\s$ and $\ce$ and $\s$ and $\cR$ are given by
$$
V_{\cal SE}\in {\mathfrak M}_{\cal S}\otimes  {\mathfrak M}_{\cal E}\ \
\mbox{and} \ \ V_{\cal SR}\in {\mathfrak M}_{\cal S}\otimes  {\mathfrak M}_{\cal R}
$$
respectively, and the total interaction is
\be
V(\lambda)=\lambda_1V_{\cal SR}+\lambda_2V_{\cal SE}\in {\mathfrak M}_{\cal S}\otimes  {\mathfrak M}_{\cal R}\otimes {\mathfrak M}_{\cal E},
\label{-1}
\ee
where $\lambda_1$, $\lambda_2$ are coupling constants ($\lambda=(\lambda_1,\lambda_2)$). The full (Schr\"odinger) dynamics is
\begin{equation}
\psi\mapsto U(m)\psi,
\label{-5}
\end{equation}
where $U(m)$ is the unitary map
\begin{equation}
U(m) = \e^{-\i\tau (L_0+V_m)}\e^{-\i \tau (L_0+V_{m-1})}\cdots \e^{-\i \tau (L_0+V_1)},
\label{-6}
\end{equation}
$\tau>0$ being the time-scale of the repeated interaction and
$V_k$ being the operator $V(\lambda)$, \fer{-1}, acting
nontrivially on $\cH_\s$, $\cH_\cR$ and the $k$-th factor
$\cH_\ce$ of $\cH_\cc$ (we will also write $L_m=L_0+V_m$). We
discuss here the dynamics \fer{-5} at discrete time steps $m\tau$
only, a discussion for arbitrary continuous times follows in a
straightforward manner by decomposing $t=m\tau +s$,
$s\in[0,\tau)$, see \cite{bjm1}.

\medskip

\noindent {\it Explicit form of finite systems and thermal reservoirs.\ }

(A) {\it Finite systems.}\ We take $\cS$ (and possibly $\cE$) to be finite. The Hamiltonian of $\cS$ is given by $H_\cs$, acting on $\fh_\cs$. The (Gelfand-Naimark-Segal) Hilbert space, the observable algebra and the Liouville operator are given by
\begin{equation*}
\cH_\cs= \fh_\cs\otimes\fh_\cs,\qquad \mm_\s ={\cal B}(\cH_\cs)\otimes\bbbone, \qquad L_\cs=H_\cs\otimes\bbbone - \bbbone\otimes H_\cs.
\label{-16}
\end{equation*}
The reference state is chosen to be the trace state, represented by
\begin{equation*}
\Psi_\s = \frac{1}{\sqrt{\dim \cH_\cs}} \sum_{j=1}^{\dim \cH_\cs}\varphi_j\otimes\varphi_j,
\label{-17}
\end{equation*}
where $\{\varphi_j\}$ is an orthonormal basis of $\fh_\s$ diagonalizing $H_\cs$.

(B) {\it Thermal reservoirs.}\
We take $\cR$ (and possibly $\cal E$) to be a thermal reservoir of free Fermi particles at a temperature $T>0$, in the thermodynamic limit. Its description was originally given in the work by Araki and Wyss \cite{Awyss}; see also \cite{JP} and \cite{MStarr},  Appendix A, for an exposition close to ours. The Hilbert space is the anti-symmetric Fock space
\begin{equation*}
\cH_\cR = \Gamma_-({\fh}):=\bigoplus_{n\geq 0} {\cal P}_-[L^2(\fh)]^{\otimes_{j=1}^n}
\label{-10}
\end{equation*}
over the one-particle space
\begin{equation}
\fh = L^2({\mathbb R},\frak G),
\label{-11}
\end{equation}
where ${\cal P}_-$ is the orthonormal projection onto the subspace of anti-symmetric functions, and $\frak G$ is an `auxiliary space' (typically an angular part like $L^2(S^2)$). In this representation, the one-particle Hamiltonian $h$ is the operator of multiplication by the radial variable (extended to negative values ) $s\in\mathbb R$ of \fer{-11},
\begin{equation*}
h =s.
\label{-12}
\end{equation*}
The Liouville operator is the second quantization of $h$,
\begin{equation}
L_\cR = \d\Gamma(h):= \bigoplus_{n\geq 0} \sum_{j=1}^n h_j,
\label{-13}
\end{equation}
where $h_j$ is understood to act as $h$ on the $j$-th factor of ${\cal P}_-[L^2(\fh)]^{\otimes_{j=1}^n}$ and trivially on the other ones.

The von Neumann algebra $\mm_\cR$ is the subalgebra of ${\cal B}(\cH_\cR)$ generated by the thermal fermionic field operators (at inverse temperature $\beta$), represented on $\cH_\cR$ by
\begin{equation*}
\varphi(g_{\beta}) = \frac{1}{\sqrt 2}\big[ a^*(g_{\beta}) +a(g_{\beta})\big].
\label{-14}
\end{equation*}
Here, we define for $g\in L^2({\mathbb R}_+,{\frak G})$
\begin{equation*}
g_\beta(s) = \sqrt{\frac{1}{\e^{-\beta s}+1}}
\left\{
\begin{array}{ll}
g(s) & \mbox{if $s\geq 0$}\\
\overline{g}(-s) & \mbox{if $s<0$.}
\end{array}
\right.
\label{-15}
\end{equation*}
We choose the reference state to be thermal equilibrium state, represented by the vacuum vector of $\cH_\cR$,
\begin{equation*}
\Psi_\cR = \Omega.
\label{-18}
\end{equation*}

\noindent We provide a precise derivation of the above formalism
starting from the usual description of a reservoir of
non-interacting and non-relativistic fermions in Section
\ref{ssec:explexa}.


\subsection{Convergence to asymptotic state}
\label{ssm1}

One of our main interests is the behaviour of averages
$\rho(U(m)^* O_mU(m))$ as $m\rightarrow\infty$, where $\rho$ is
any (normal) initial state of the total system, and where $O_m$ is
a so-called {\it instantaneous observable} \cite{bjm1, bjm2,
bjm3}. Let $A_{\cs\cR}\in\mm_\cs\otimes\mm_\cR$ and let
$B_j\in\mm_\ce$, $j=-l,\ldots,r$, where $l,r\geq 0$ are integers.
The associated instantaneous observable is
\begin{equation}
O_m=A_{\cs\cR} \otimes_{j=m-l}^{m+r}\vartheta_j(B_{j-m}) \in \mm
\label{-7}
\end{equation}
where $\vartheta_j(B)$ is the observable of $\mm$ which acts as
$B$ on the $j$-th factor of $\cH_\cc$, and trivially everywhere
else ($\vartheta_j$ is the translation to the $j$-th factor). An
instantaneous observable is a time-dependent one. It may be viewed
as a train of fixed observables moving with time along the chain
$\cC$ so that at time $m$ it is ``centered'' at the $m$-th factor
$\cH_\ce$ of $\cH_\cc$, on which it acts as $B_0$. If $O$ acts
trivially on the elements of the chain, then the corresponding
instantaneous observable is constant and $O_m=O$. However, in
order to be able to reveal interesting physical properties of the
system, instantaneous observables are needed. For instance
observables measuring fluxes of physical quantities (like energy,
entropy) between $\s$ and the chain involve instantaneous
observables acting non-trivially on $\ce_{m}$ and on $\ce_{m+1}$,
which corresponds to nontrivial $A_{\cs\cR}$ and $B_0$, $B_1$. We
denote the Heisenberg dynamics of observables by (see \fer{-6})
\begin{equation}
\alpha^m(O_m) = U(m)^* O_m U(m).
\label{-6.5}
\end{equation}

The total reference vector
\begin{equation}
\Psi_0=\Psi_\cs\otimes\Psi_\cR\otimes\Psi_\cC
\label{-23}
\end{equation}
is cyclic and separating for the von Neumann algebra $\mm$. We also introduce, for later purposes, the projector $P_{\cal SR}=\un_{\cal SR}\otimes |\Psi_{\cal C}\ket\bra \Psi_{\cal C}|$, which range we often identify with ${\cH}_{\cal SR}=\cH_\cS\otimes \cH_\cR$. Let $J$ and $\Delta$ be the modular conjugation and the modular operator associated to the pair $(\mm,\Psi_0)$ \cite{BR}. In order to represent the dynamics in a convenient way (using a so-called $C$-Liouville operator), we make the following assumption.

\begin{itemize}
\item[{\bf H1}] The interaction operator $V(\lambda)$, \fer{-1}, satisfies
$\ds \Delta^{1/2}V(\lambda)\Delta^{-1/2}\in {\mathfrak M}_{\cal S}\otimes {\mathfrak M}_{\cal R}\otimes {\mathfrak M}_{\cal E}.$
\end{itemize}

Since we will be using analytic spectral deformation methods on the factor $\cH_\cR$ of $\cH$, we need to make a regularity assumption on the interaction. Let $\R\ni \theta\mapsto T(\theta)\in \cB(\cH_\cR)$ be the unitary group defined by
\be
T(\theta)=\Gamma(e^{-\theta \partial_s}) \ \ \mbox{on} \ \ \Gamma_-(L^2(\R,{\mathfrak G})),
\label{mm9}
\ee
where for any $f\in L^2(\R,{\mathfrak G})$,
$$
(e^{-\theta \partial_s}f)(s)=f(s-\theta).
$$
In the following, we will use the notation
$$
T(\theta)=\I_\cS\otimes T(\theta)\otimes \I_\cE
$$
for simplicity. Note that $T(\theta)$ commutes with all observables acting trivially on $\cH_\cR$, in particular with $P_{\cal SR}$. Also, we have $T(\theta)\Psi_\cR=\Psi_\cR$ for all $\theta$. The spectral deformation technique relies on making the parameter $\theta$ complex.
\begin{itemize}
\item[{\bf H2}] The coupling operator $W_{\cS\cR}:= V_{\cs\cR}
-J\Delta^{1/2} V_{\cs\cR}\Delta^{-1/2}J$ is translation analytic
in a strip $\kappa_{\theta_0}=\{ z\ :\ 0<\Im z<\theta_0\}$ and
strongly continuous on the real axis. More precisely, there is a
$\theta_0>0$ such that the map
$$
{\mathbb R}\ni\theta \mapsto T^{-1}(\theta) W_{\cal SR}T(\theta)= W_{\cal SR}(\theta)\in \fm_\cS\otimes\fm_\cR,
$$
admits an analytic continuation into $\theta\in\kappa_{\theta_0}$
which is strongly continuous as $\Im \theta\downarrow 0$, and
which satisfies
$$
\sup_{0\leq \Im\theta<\theta_0} \| W_{\cS\cR}(\theta)\| < \infty.
\label{mm1}
$$
\end{itemize}
Let $O_m$ be an instantaneous observable \fer{-7}. We say that
$O_m$ is an {\it analytic observable} if
\begin{equation}
T(\theta)^{-1} O_m \Psi_0
\label{-22}
\end{equation}
has an analytic extension to $\theta\in\kappa_{\theta_0}$ which is
continuous on the real axis. Evidently, since $T$ acts on
$\cH_\cR$ only, this is equivalent with
$T(\theta)^{-1}A_{\cs\cR}\Psi_0$ having such an extension.

Finally we present a `Fermi golden rule condition' which
guarantees that the subsystems are well coupled so that the
physical phenomena studied are visible at lowest nontrivial order
in the pertrubation $\lambda_1, \lambda_2$. This is a very common
hypothesis which is most often verified in concrete applications.
To state it, we mention that the evolution is generated by
so-called {\it reduced dynamics operators} \cite{bjm1}
$M(\lambda)$ acting on the reduced space $\cH_{\cs\cR}$ (where the
degrees of freedom of $\cc$ have been `traced out'). In this
paper, we analyze the spectrally deformed operators
\begin{equation*}
M_\theta(\lambda) = T(\theta)^{-1}M(\lambda)T(\theta).
\label{-21}
\end{equation*}
We show in Corollary \ref{mmcor1} that $M_\theta(\lambda)$ has an
analytic extension into the strip $\kappa_{\theta_0}$, in the
sense of H2 above. We show that $1$ is an eigenvalue of
$M_\theta(\lambda)$ for all $\theta,\lambda$ and that, for small
couplings $\lambda=(\lambda_1, \lambda_2)$ ,  the spectrum of
$M_\theta(\lambda)$ must lie in the closed unit disk. The latter
fact is true because $M_\theta(\lambda)$ is the analytically
translated (one-step) propagator of a reduced unitary dynamics,
although we can only prove it in a perturbative regime. The former
fact can be seen as a normalization (the trace of the reduced
density matrix of $\cs+\cR$ equals unity at all times). Since the
propagator at time step $m$ is represented by a power of
$M_\theta(\lambda)$, it is not surprising that convergence to a
final state is related to the peripheral eigenvalues of
$M_\theta(\lambda)$. The following Fermi golden rule condition is
an ergodicity condition ensuring the existence of a unique limit
state.
\begin{itemize}
\item[{\bf FGR}] There is a $\theta_1\in\kappa_{\theta_0}$ and a
$\lambda_0>0$ (depending on $\theta_1$ in general) such that for
all $\lambda$ with $0<|\lambda|< \lambda_0$,
$\sigma(M_{\theta_1}(\lambda))$ (spectrum) lies inside the complex
unit disk, and $\sigma(M_{\theta_1}(\lambda))\cap {\mathbb
S}=\{1\}$, the eigenvalue 1 being simple and isolated. (${\mathbb
S}$ is the complex unit circle.)
\end{itemize}

This condition is verified in practice by perturbation theory
(small $\lambda$). It is also possible to prove that if the
spectral radius of $M_{\theta_1}(\lambda)$ is determined by
discrete eigenvalues only, then the spectrum of
$M_{\theta_1}(\lambda)$ is {\it automatically} inside the unit
disk (see Proposition \ref{draco}). Since the spectrum is a closed
set the FGR condition implies that apart from the eigenvalue 1 the
spectrum is contained in a disk of radius $\e^{-\gamma}<1$.

\begin{thm}[Convergence to asymptotic state]
\label{1st}
Assume that assumptions H1, H2 and FGR are satisfied. Then there is a $\lambda_0>0$ s.t. if $0<|\lambda|<\lambda_0$, the following holds. Let $\rho$ be any normal initial state on $\frak M$, and let $O_m$ be an analytic instantaneous observable of the form \fer{-7}. Then
\begin{equation}
\lim_{m\ra\infty} \rho\big (\alpha^m(O_m)\big) =
\rho_{+,\lambda}\Big(P_{\cS\cR}\
\alpha^{l+1}\big(A_{\cs\cR}\otimes_{j=-l}^0B_j\big)
P_{\cS\cR}\Big) \prod_{j=1}^r \langle\Psi_\ce |  B_j\Psi_\ce
\rangle, \label{mm13}
\end{equation}
where $\rho_{+,\lambda}$ is a state on $\mm_\cS\otimes\mm_\cR$, $P_{\cS\cR}$ is the orthogonal projection onto the subspace $\cH_\cs\otimes\cH_\cR$, and where $\alpha^l$ is the dynamics \fer{-6.5}. Moreover, for analytic $A\in\mm$, we have the representation
\begin{equation}
\rho_{+,\lambda}(P_{\cS\cR}AP_{\cS\cR})=\bra\psi^*_{\theta_1}(\lambda)| T(\theta_1)^{-1} P_{\cS\cR} A P_{\cS\cR} \Psi_\cS\otimes\Psi_\cR\ket,
\label{-9}
\end{equation}
where $\psi_{\theta_1}^*(\lambda)$ is the unique invariant vector of the adjoint operator $[M_{\theta_1}(\lambda)]^*$, normalized as $\langle \psi_{\theta_1}^*(\lambda) | \Psi_0\rangle=1$.
\end{thm}
{\it Remark.\ } The operators $B_j$ with $j\geq 1$ measure
quantities on elements $\ce_{m+j}$ which, at time $m$, have not
yet interacted with the system $\cs$. Therefore they evolve
independently simply under the evolution of $\cE_{m+j}$. For large
times $m\rightarrow\infty$, the elements of the chain approach the
reference state $\Psi_\cE$ (because the initial state is normal),
and the latter is stationary w.r.t. the uncoupled evolution. This
explains the factorization in \fer{mm13}.

As a special case of Theorem \ref{1st} we obtain the reduced evolution of $\cs+\cR$.
\begin{cor}
\label{mcor1}
Assume the setting of Theorem \ref{1st}. Then
\begin{equation*}
\lim_{m\ra\infty} \rho\big(\alpha^m(A_{\cs\cR})\big) = \rho_{+,\lambda}\big( A_{\cs\cR}\big).
\label{-30}
\end{equation*}
\end{cor}


\subsection{Thermodynamic properties of asymptotic state}
\label{sectthermo}

The total energy of the system is not defined, since $\cR$ and $\cc$ are reservoirs (and typically have infinite total energy). However, the {\it energy variation} is well defined. More precisely, the formal expression for the energy at time $m$, $\alpha^m(L_0+\vartheta_m(V))$ (see \fer{-3}, \fer{-1}), undergoes a jump
\begin{eqnarray*}
\Delta E^{\mbox{\scriptsize tot}}(m)&=&\alpha^{m+1}(L_0+\vartheta_{m+1}(V))-\alpha^m(L_0+\vartheta_m(V))\nonumber\\
&=& \lambda_2 \alpha^m(\vartheta_{m+1}(V_{\cs\ce})-\vartheta_m(V_{\cs\ce})),
\label{-50}
\end{eqnarray*}
as time passes the moment $m\tau$. The variation $\Delta E^{\mbox{\scriptsize tot}}(m)$ is thus an instantaneous observable. In applications this observable is analytic and hence we obtain under the conditions of Theorem \ref{1st} (see also \cite{bjm1}) that
\begin{equation*}
\d E_+^{\rm tot}:= \lim_{m\rightarrow\infty} \frac1m\rho( \Delta E^{\mbox{\scriptsize tot}}(m) ) = \rho_{+,\lambda}(j_+^{\rm tot}),
\label{-52}
\end{equation*}
where $j_+^{\rm tot} = V-\alpha^\tau(V)$ is the total energy flux observable.
The quantity $\d E_+^{\rm tot}$ represents the asymptotic energy change per unit time $\tau$ of the entire system. In the same way we define the variation of energy within the system $\cs$, the reservoir $\cR$ and the chain $\cc$ between times $m$ and $m+1$ by
\begin{eqnarray*}
\Delta E^{\cs}(m) &=& \alpha^{m+1}(L_\cs)-\alpha^m(L_\cs),\label{-53.1}\\
\Delta E^{\cR}(m) &=& \alpha^{m+1}(L_\cR)-\alpha^m(L_\cR),\label{-53}\\
\Delta E^{\cC}(m) &=& \alpha^{m+1}(L_{\cE_{m+1}})-\alpha^m(L_{\cE_{m+1}}).
\end{eqnarray*}
These variations can be expressed in terms of commutators
$[V_{\cs\ce},L_\#]$ and $[V_{\cs\cR},L_\#]$, where $\#=\cs, \ce,
\cR$. Since $[V_{\cs\ce},L_\#]$ acts on ${\cal S}+{\cal E}$ only,
it is an analytic observable (see sentence after \fer{-22}). We
make the following Assumption.
\begin{itemize}
\item[{\bf H3}] The commutators $[V_{\cs\cR},L_\#]$, where
$\#=\cs, \ce, \cR$, are analytic observables in $\mm$.
\end{itemize}
We can thus apply Theorem \ref{1st} to obtain (see Section \ref{sectvarenergy})
\begin{equation}
\d E^\#_+ :=\lim_{m\rightarrow\infty} \frac1m\rho(\Delta E^\#(m)) = \rho_{+,\lambda}(j_+^\#),\qquad \#=\cs, \cR, \cc,
\label{-55}
\end{equation}
where $j^\#$ are explicit `flux observables' (c.f. \fer{energyobs}-\fer{enobs2}). We show in Proposition \ref{fluxprop} that $j_+^{\rm tot} =  j_+^{\cs} +j_+^{\cR}+ j_+^{\cc}$, and that $\rho_{+,\lambda}(j_+^\cs)=0$. It follows immediately that
\begin{equation}
\d E_+^{\rm tot} = \d E_+^\cR +\d E_+^\cc.
\label{-56}
\end{equation}
The total energy variation is thus the sum of the variations in the energy of $\cC$ and $\cR$. The details of how the energy variations are shared between the subsystems depends on the particulars of the model considered; see below for an explicit example.\\

\medskip

Next we consider the {\it entropy production}. Given two normal
states $\rho$ and $\rho_0$ on $\fm$, the relative entropy of
$\rho$ with respect to $\rho_0$ is denoted by ${\rm
Ent}(\rho|\rho_0)$. (This definition coincides with the one in
\cite{bjm1} and differs from certain other works by a sign; here
${\rm Ent}(\rho|\rho_0)\geq 0$).

We examine the change of relative entropy of the state of the system as time evolves, relative to the reference state $\rho_0$ represented by the reference vector $\Psi_0$, see \fer{-23}. For a thermodynamic interpretation of the entropy, we take the vectors $\Psi_\#$, $\#=\cS,\, \cE,\, \cR$ to represent equilibrium states of respective temperatures $\beta_\cs$, $\beta_\ce$, $\beta_\cR$. We analyze the change of relative entropy
$$
\Delta S(m)={\rm Ent}(\rho\circ \alpha^m|\rho_0)-{\rm Ent}(\rho|\rho_0)
$$
proceeding as in \cite{bjm1}. We show in Section \ref{sectvarenergy} (see \fer{eq:entropyprod}) that
$$
\d S_+:= \lim_{m\to\infty} \frac{\Delta S(m)}{m}= (\beta_\cR-\beta_\cE) \d E_+^\cR + \beta_\cE \d E^{{\rm tot}}_+.
$$
Combining this result with \fer{-56}, we arrive at
\begin{cor}
\label{mcor2}
The system satisfies the following \emph{asymptotic 2nd law of thermodynamics},
$$
\d S_+= \beta_\cE \d E_+^\cC+\beta_\cR \d E_+^\cR.
$$
\end{cor}


\subsection{An explicit example}
\label{ssec:explexa}

We consider $\cs$ and $\ce$ to be two-level systems. The
observable algebra for $\cS$ and for $\cE$ is
$\fa_\cS=\fa_\cE=M_2(\C)$. Let $E_\cS$, $E_\cE>0$ be the
``excited'' energy level of $\cS$ and of $\cE$, respectively.
Accordingly, the Hamiltonians are given by
$$
h_\cS= \left( \begin{array}{cc} 0 & 0 \\
0 & E_\cS \end{array} \right)
\mbox{\ \ and \ \ }
h_\cE= \left( \begin{array}{cc} 0 & 0 \\
0 & E_\cE \end{array} \right).
$$
The dynamics are $\alpha_\cS^t(A)= \e^{\i t h_\cS}A\e^{-\i t h_\cS}$  and  $\alpha_\cE^t(A)= \e^{\i t h_\cE}A\e^{-\i t h_\cE}$.
We choose the reference state of $\cE$ to be the Gibbs state at inverse temperature $\beta_\cE$, i.e.
\begin{equation*}
\rho_{\beta_\cE,\cE}(A)=\frac{\tr(\e^{-\beta_\cE h_\cE}A)}{Z_{\beta_\cE,\cE}}, \ \ {\rm where} \ \   Z_{\beta_\cE,\cE}=\tr(\e^{-\beta_\cE h_\cE}),
\label{mm110}
\end{equation*}
and we choose (for computational convenience) the reference state for $\cS$ to be the tracial state, $\rho_{\cS}(A)=\frac{1}{2}\tr(A)$. The interaction operator between $\cS$ and an element $\cE$ of the chain is defined by $\lambda_2 v_{\cS\cE}$, where $\lambda_2$ is a coupling constant, and
\begin{equation*}\label{spinspininter}
v_{\cS\cE}:=a_\cS\otimes a_\cE^*+a_\cS^*\otimes a_\cE.
\end{equation*}
The above creation and annihilation operators are represented by the matrices
$$
a_{\#}= \left( \begin{array}{cc} 0 & 1 \\
0 & 0 \end{array} \right)
\mbox{\ \ and\ \ }
a_{\#}^*= \left( \begin{array}{cc} 0 & 0 \\
1 & 0 \end{array} \right).
$$
To get a Hilbert space description of the system, one performs the Gelfand-Naimark-Segal (GNS) construction of $(\fa_\cS,\rho_{\cS})$ and $(\fa_\cE,\rho_{\beta_\cE,\cE})$, see e.g. \cite{BR,bjm1}. In this representation, the Hilbert spaces are given by
$$
\cH_\cS = \cH_\cE = {\mathbb C}^2\otimes{\mathbb C}^2,
$$
the Von Neumann algebras by
$$
\fm_\cS=\fm_\cE=M_2(\C)\otimes \one_{\C^2}\subset \cB(\C^2\otimes\C^2),
$$
and the vectors representing $\rho_{\cS}$ and $\rho_{\beta_\cE,\cE}$ are
$$
\Psi_\cS = \frac{1}{\sqrt 2} \left(|0\ket\otimes|0\ket+|1\ket\otimes|1\ket\right),\qquad \Psi_\cE=\frac{1}{\sqrt{{\rm Tr}\, \e^{-\beta_\cE h_\cE}}}\left( |0\ket\otimes|0\ket + \e^{-\beta_\cE E_\cE/2}|1\ket\otimes|1\ket\right).
$$
In other words,
$\rho_\cS(A)=\scalprod{\psi_{\cS}}{(A\otimes\bbbone)\psi_{\cS}}$ and $\rho_{\beta_{\cE},\cE}(A)=\scalprod{\psi_{\cE}}{(A\otimes\bbbone)\psi_{\cE}}$. Above,
$|0\ket$ (resp. $|1\ket$) denotes the ground (resp. excited) state of
$h_\cS$ and $h_\cE$. For shortness, in the following we will denote $|ij\ket$ for $|i\ket\otimes|j\ket$, $i,j=0,1$. The free Liouvilleans $L_\cS$ and $L_\cE$ are given by
$$
L_\cS = h_\cS\otimes \one_{\C^2}-\one_{\C^2}\otimes h_\cS , \quad L_\cE=h_\cE\otimes \one_{\C^2}-\one_{\C^2}\otimes h_\cE
$$
and the interaction operator $V_{\cS\cE}$ is
$$
V_{\cS\cE}=(a_\cS\otimes\one_{\C^2})\otimes(a^*_\cE\otimes
 \one_{\C^2}) +(a_\cS^*\otimes\one_{\C^2})\otimes(a_\cE\otimes \one_{\C^2}).
$$

\medskip

For the reservoir, we consider a bath of non-interacting and non-relativistic fermions. The one particle space is $\fh_\cR=L^2(\R^3, \d^3 k)$ and the one-particle energy operator $h_\cR$ is the multiplication operator by $|k|^2$. The Hilbert space for the reservoir is thus $\Gamma_-(\fh_\cR)$ and the Hamiltonian is the second quantization $\d\Gamma(h_\cR)$ of $h_\cR$ (see \fer{-13}). The algebra of observables is the $C^*$-algebra of operators
$\mathfrak A$ generated by $\{a^\#(f)\, |\, f\in\fh_\cR\}$ where $a/a^*$ denote the usual annihilation/creation operators on $\Gamma_-(\fh_\cR)$. The dynamics is
 given by $\tau_{\rm f}^t(a^\#(f))=a^\#(\e^{\i th}f)$,
where $h$ is the Hamiltonian of a single particle, acting on
$\fh$. It is well known (see e.g. \cite{BR}) that for any
$\beta_\cR>0$ there is a unique $(\tau_{\rm f},\beta)-$KMS state
$\rho_{\beta_\cR}$ on $\mathfrak A$, determined by the
two point function $\rho_{\beta_\cR}(a^*(f)a(f))=\langle f,
(1+\e^{\beta_\cR h_\cR})^{-1}f\rangle$, and which we choose to be the reference state of the reservoir. Finally, the interaction between the small system $\cS$ and the reservoir is chosen of electric dipole type, i.e. of the form $v_{\cs\cR}=(a_\cS+a_\cS^*)\otimes \varphi_\cR(f)$ where $f\in \fh_\cR$ is a form factor and $\varphi(f)=\frac{1}{\sqrt{2}}(a(f)+a^*(f))$.

We know explain how to get a descrpition of the reservoir similar to the one given in Section \ref{sec:descripintro}. As for $\cS$ and $\cE$, the first point is to perform the GNS representation of $(\mathfrak A, \rho_{\beta_\cR})$, so called Araki-Wyss representation \cite{Awyss}. Namely, if $\Omega$ denotes the Fock vacuum and $N$ the number operator of $\Gamma_-(\fh_\cR)$, the Hilbert space is given by
\begin{equation*}\label{eq:awspace1}
\tilde{\cH}_\cR= \Gamma_-(L^2(\R^3, \d^3 k)) \otimes \Gamma_-(L^2(\R^3, \d^3 k)),
\end{equation*}
the Von-Neumann algebra of observables is
\begin{equation*}
\tilde{\fm}_\cR = \pi_\beta\left( \mathfrak A\right)''
\end{equation*}
where
\begin{equation*}
\begin{array}{l}
\pi_{\beta}(a(f)) = a\left(\frac{\e^{\beta h/2}}{\sqrt{1+\e^{\beta
 h}}}f \right)\otimes \one+(-1)^N\otimes a^*\left(\frac{1}{\sqrt{1+
 \e^{\beta h}}}\bar{f} \right)=:a_{\beta}(f),\\
\pi_{\beta}(a^*(f)) = a^*\left(\frac{\e^{\beta
h/2}}{\sqrt{1+\e^{\beta h}}}f \right)\otimes
\one+(-1)^N\otimes a\left(\frac{1}{\sqrt{1+\e^{\beta
h}}}\bar{f} \right)=:a^*_{\beta}(f),
\end{array}
\end{equation*}
the reference vector is
\begin{equation*}
\tilde{\Psi}_\cR=\Omega \otimes \Omega,
\end{equation*}
and the Liouvillean is
\begin{equation*}
\tilde{L}_\cR=\d\Gamma(h_\cR)\otimes\one -\one\otimes \d\Gamma(h_\cR).
\end{equation*}
We then consider the isomorphism between $L^2(\R^3,\d^3k)$ and $L^2(\R^+\times S^2, \frac{\sqrt{r}}{2} \d r \d\sigma)\simeq L^2(\R^+,\frac{\sqrt{r}}{2} \d r; \frak G)$, where ${\frak G}=L^2(S^2,\d\sigma)$, so that the operator $h_\cR$ (the multiplication by $|k|^2$) becomes multiplication by $r\in \R^+$ (i.e. we have $r=|k|^2$). The Hilbert space $\tilde{\cH}_\cR$ is thus isomorphic to
\begin{equation}\label{eq:awspace2}
\Gamma_-\left( L^2(\R^+,\frac{\sqrt{r}}{2} \d r; \frak G)\right)\otimes \Gamma_-\left( L^2(\R^+,\frac{\sqrt{r}}{2} \d r; \frak G)\right).
\end{equation}
Next we make use of the maps
$$
a^\#(f)\otimes \one \mapsto a^\#(f\oplus 0),\quad (-1)^N \otimes a^\#(f)\mapsto a^\#(0\oplus f)
$$
to define an isometric isomorphism between (\ref{eq:awspace2}) and
\begin{equation*}\label{eq:awspace3}
\Gamma_-\left( L^2(\R^+,\frac{\sqrt{r}}{2} \d r; {\frak G}) \oplus L^2(\R^+,\frac{\sqrt{r}}{2} \d r; \frak G) \right).
\end{equation*}
A last isometric isomorphism between the above Hilbert space and
\begin{equation*}\label{eq:reservspace}
\cH_\cR:= \Gamma_-\left( L^2(\R,\d s; {\frak G})\right)
\end{equation*}
is induced by the following isomorphism between the one-particle spaces $L^2(\R^+,\frac{\sqrt{r}}{2} \d r; {\frak G}) \oplus L^2(\R^+,\frac{\sqrt{r}}{2} \d r; \frak G)$ and $L^2(\R,\d s; {\frak G})=:\fh$
\begin{equation*}
f\oplus g \mapsto h, \ {\rm where} \ h(s)=\frac{|s|^{1/4}}{\sqrt{2}}\left\{ \begin{array}{ll} f(s) & {\rm if} \ s\geq 0,\\ g(-s) & {\rm if} s<0.  \end{array}\right.
\end{equation*}

Using the above isomorphisms, one gets a descrition of the form given in Section \ref{sec:descripintro} for the reservoir $\cR$. In this representation, the interaction operator $v_{\cs\cR}$ becomes
\begin{equation*}
V_{\cS\cR}=(\sigma_x\otimes \one_{\C^2}) \otimes  \varphi(f_{\beta_\cR}) \in \fm_\cS\otimes \fm_\cR,
\end{equation*}
where $\sigma_x=a_\cS+a_\cS^*$ is the Pauli matrix and $f_{\beta_\cR}\in \fh=L^2(\R,\d s; L^2(S^2,\d \sigma))$ is related to the initial form factor $f\in L^2(\R^3,\d^3k)$ as follows
\begin{equation}\label{eq:awformfactor}
\left(f_{\beta_\cR}(s)\right)(\sigma)=\frac{1}{\sqrt{2}}\frac{|s|^{1/4}}{\sqrt{1+\e^{-\beta_\cR s}}}\left\{  \begin{array}{ll} f(\sqrt{s}\;\sigma) & {\rm if} \ s\geq 0, \\ \bar{f}(\sqrt{-s}\;\sigma)& {\rm if} \ s< 0. \end{array}\right.
\end{equation}

\medskip
As mentioned at the beginning of the introduction the situation where $\cs$ is interacting with $\cR$ or $\cc$ alone has been treated in previous works \cite{JP, MStarr} and \cite{bjm1}. If $\cs$ is coupled to $\cR$ alone, then a normal initial state approaches the joint equilibrium state, i.e. the equilibrium state of the coupled system $\cS+\cR$ at temperture $\beta_\cR^{-1}$, with speed $\e^{-m\tau\gamma_{\rm th}}$ (we consider discrete moments in time, $t=m\tau$ to compare with the repeated interaction situation). If $\cs$ is coupled to $\cc$ alone, initial normal states approach a repeated interaction asymptotic state, which turns out to be the equilibrium state of $\cS$ at inverse temperature $\beta'_\ce$ where
\begin{equation}
\beta'_\ce = \beta_\ce \frac{E_\ce}{E_\cs},
\label{betaprime}
\end{equation}
and with speed $\e^{-m\tau\gamma_{\rm ri}}$. The convergence rates are given by
\begin{eqnarray}
\gamma_{\rm th}&=& \lambda_1^2\gamma_{\rm th}^{(2)}+O(\lambda_1^4),\qquad \mbox{with}  \quad \gamma_{\rm th}^{(2)}=\frac{\pi}{2} \sqrt{E_\cs}\|f(\sqrt{E_\cS})\|_{\mathfrak G}^2 \label{thermrate}\\
\gamma_{\rm ri}&=& \lambda_2^2\gamma_{\rm ri}^{(2)}+O(\lambda_2^4), \qquad \mbox{with} \quad \gamma_{\rm ri}^{(2)}=\tau\sinc^2\left( \frac{\tau(E_\cE-E_\cS)}{2}\right),\label{rirate}
\end{eqnarray}
where $\sinc(x)=\sin(x)/x$ and $\ds \|f(\sqrt{E_\cS})\|_{\mathfrak G}^2:= \int_{S^2} |f(\sqrt{E_\cS}\; \sigma)|^2\d\sigma$.

\medskip

In order to satisfy the translation analyticity requirement {\bf H2}, we need to make some assumption on the form factor $f$.
Let $I(\delta)\equiv \{z\in\C,\, |\Im(z)|<\delta\}$. We denote by $H^2(\delta)$ the Hardy class of analytic functions $h: I(\delta)\to {\mathfrak G}$ which satisfy
$$
\|h\|_{H^2(\delta)}:= \sup_{|\theta|<\delta} \ \int_\R \|h(s+i\theta)\|_{\mathfrak G}^2 \d s < \infty.
$$

\begin{itemize}
\item[{\bf H4}] Let $f_0$ be defined by \fer{eq:awformfactor}, with $\beta_\cR=0$. There is a $\delta>0$ s.t. $\e^{-\beta_\cR s/2} f_0(s)\in H^2(\delta)$.
\end{itemize}

\begin{prop}[Asymptotic state of $\cs$]
\label{ssf:asymptstate}
Assume $f$ satisfies {\bf H4},  $\|f(\sqrt{E_\cS})\|_{\mathfrak G}\neq 0$ and $\tau(E_\cE-E_\cS)\notin 2\pi \Z^*$. Then the asymptotic state $\rho_{+,\lambda}$ is given by
$$
\rho_{+,\lambda}= \left( \gamma \rho_{\beta_\cR,\cS}+ (1-\gamma) \rho_{\beta_\cE',\cS} \right)\otimes \rho_{\beta_\cR,\cR} +O(\lambda),
$$
where $\rho_{\beta,\#}$ is the Gibbs state of $\#$, $\#=\cS,\cR$, at inverse temperature $\beta$ and where $\gamma$ is given by
\begin{equation*}
\gamma =  \frac{\lambda_1^2\gamma_{\rm th}^{(2)}}{\lambda_1^2\gamma_{\rm th}^{(2)} + \lambda_2^2\gamma_{\rm ri}^{(2)}}. \label{mm-59}
\end{equation*}
\end{prop}

\noindent {\bf Remark.} The fact that the asymptotic state
$\rho_{+,\lambda}$ is a convex combination of the two asymptotic
states $\rho_{\beta_\cR,\cS}$ and $\rho_{\beta_\cE',\cS}$ holds
only because the system $\cS$ is a two-level system and is not
true in general.

\medskip

Using \fer{-59}-\fer{enobs2}, Corollary \ref{mcor2} and Proposition \ref{ssf:asymptstate}, an explicit calculation of the energy fluxes and the entropy production for this concrete model reveals the following result.
\begin{prop}
\label{ssf:fluxes}
Assume that $\|f(\sqrt{E_\cS})\|_{\mathfrak G}\neq 0$ and $\tau(E_\cE-E_\cS)\notin 2\pi \Z^*$. Then
\begin{eqnarray*}
\d E_+^\cC & = & \kappa E_\cE \left( \e^{-\beta_\cR E_\cS}- \e^{-\beta_\cE' E_\cS} \right)+O(\lambda^3),\\
\d E_+^\cR & = & \kappa E_\cS \left( \e^{-\beta_\cE' E_\cS}- \e^{-\beta_\cR E_\cS} \right)+O(\lambda^3),\\
\d E_+^{{\rm tot}} & = & \kappa (E_\cE-E_\cS) \left( \e^{-\beta_\cR E_\cS}- \e^{-\beta_\cE' E_\cS} \right)+O(\lambda^3),\\
\d S_+ & = & \kappa (\beta_\cE' E_\cS-\beta_\cR E_\cS) \left( \e^{-\beta_\cR E_\cS}- \e^{-\beta_\cE' E_\cS} \right)+O(\lambda^3),
\end{eqnarray*}
where
$$
\kappa=Z_{\beta_\cR,\cS}^{-1}Z_{\beta_\cE',\cS}^{-1}\frac{\lambda_1^2 \gamma_{\rm th}^{(2)} \ \lambda_2^2 \gamma_{\rm ri}^{(2)} }{\lambda_1^2 \gamma_{\rm th}^{(2)}+\lambda_2^2 \gamma_{\rm ri}^{(2)} }.
$$
\end{prop}

\noindent {\bf Remarks.} 1. The constant $\kappa$ is positive and of order $\lambda^2$. Moreover it is zero if at least one of the two coupling constants vanishes (we are then in an equilibrium situation and there is no energy flux neither entropy production).

2. The energy flux $\d E_+^\cC$ is positive (energy flows {\it into} chain) if and only if the reservoir temperature $T_\cR=\beta_\cR^{-1}$ is greater than the renormalized temperature $T_\cE'=\beta_\cE'^{-1}$ of the chain, i.e. if and only if the reservoir is ``hotter''. A similar statement holds for the energy flux $\d E_+^\cR$ of the reservoir. Note that it is not the temperature of the chain which plays a role but its renormalized value \fer{betaprime}.

3. When both the reservoir and the chain are coupled to the system $\cS$ ($\lambda_1 \lambda_2\neq 0$) the entropy production vanishes (at the main order) if and only if the two temperatures $T_\cR$ and $T_\cE'$ are equal, i.e. if and only if we are in an equilibrium situation. Once again, it is not the initial temperature of the chain which plays a role but the renormalized one.

4. The total energy variation can be either positive or negative depending on the parameters of the model. This is different from the situation considered in \cite{bjm1}, where that variation was always non-negative.


\section{Proof of Theorem \ref{1st}}


\subsection{Generator of dynamics $K_m$}
\label{sectK}

We recall the definition of the so-called `C-Liouvillean' introduced for the study of open systems out of equilibrium in \cite{jp2002}, and further developped in \cite{bjm1, bjm2, bjm3, msb1, msb2, MStarr} (see also references in the latter papers). Let $(J_\#, \Delta_\#)$ denote the modular data associated with $({\mathfrak M}_\#, \Psi_\#)$, with $\#$ given by $\cal S, R$ or $\cal E$. Then
$$
(J,\Delta)=(J_{\cal S}\otimes J_{\cal R}\otimes J_{\cal E}, \Delta_{\cal S}\otimes \Delta_{\cal R}\otimes \Delta_{\cal E})
$$
are the modular data associated with $({\mathfrak M}_{\cal S}\otimes {\mathfrak M}_{\cal R}\otimes {\mathfrak M}_{\cal E}, \Psi_{\cal S}\otimes \Psi_{\cal R}\otimes \Psi_{\cal E})$. We will write $J_m$ and $\Delta_m$ to mean that these operators are considered on the $m$-th copy of the infinite tensor product ${\cal H}_{\cal C}$.

We define the C-Liouville operator
$$
K_m=L_m- J_m\Delta_m^{1/2}V_m(\lambda)\Delta_m^{-1/2}J_m\equiv
L_0+W_m(\lambda),
$$
$m\geq 1$, where $W_m(\lambda)\in {\mathfrak M}_{\cal S}\otimes {\mathfrak M}_{\cal R}\otimes {\mathfrak M}_{\cal E}$ is given by
\begin{eqnarray}\label{couk}
W_m(\lambda)&=&\lambda_1(V_{\cS\cR}-(J_{\cS}\Delta^{1/2}_{\cS}\otimes J_{\cR}\Delta^{1/2}_{\cR})V_{\cS\cR}(J_{\cS}\Delta^{1/2}_{\cS}\otimes J_{\cR}\Delta^{1/2}_{\cR}))\nonumber\\
&&+\lambda_2(V_{\cS\cE, m}-(J_{\cS}\Delta^{1/2}_{\cS}\otimes J_m\Delta^{1/2}_m)V_{\cS\cE, m}(J_{\cS}\Delta^{1/2}_{\cS}\otimes J_m\Delta^{1/2}_m))\nonumber\\
&\equiv& \lambda_1W_{\cS\cR}+ \lambda_2W_{\cS\cE, m}.
\label{mm20}
\end{eqnarray}
Of course, $W_{\cs\ce, m}$ is the operator acting as $W_{\cs\ce}$ on the subspace $\cH_\cs\otimes\cH_{\ce_m}$ of $\cH$, and trivially on its orthogonal complement.

The operators $K_m$ have two crucial properties \cite{jp2002, bjm1, msb1}. The first one is that they implement the same dynamics as the $L_m$:
\begin{equation*}
\e^{\i t L_m} A \e^{-\i t L_m}=\e^{\i t K_m} A \e^{-\i t K_m},\qquad \forall t\geq 0, \forall A\in\mm_\cs\otimes\mm_\cR\otimes\mm_{\cc}.
\label{kz'}
\end{equation*}
The second crucial property is that the reference state $\Psi_0$, \fer{-23}, is left invariant under the evolution $\e^{\i t K_m}$,
\begin{equation}
K_m\Psi_0=0, \qquad \forall m.
\label{kz}
\end{equation}


\subsection{Reduced Dynamics Operator}
\label{rdo}

We follow the strategy of \cite{bjm1} to reduce the problem to the study of the high powers of an effective dynamics operator. The main difference w.r.t. \cite{bjm1} is that in the present setup, the effective dynamics operator acts now on the {\it infinite dimensional} Hilbert space ${\cal H}_{\cal S}\otimes {\cal H}_{\cal R}$.

We first split off the free dynamics of elements not interacting with $\cs$ by writing the product of exponentials in $U(m)$, \fer{-6},  as
$$
U(m)=U_m^-e^{-\i\tau L_m}e^{-\i\tau L_{m-1}}\cdots e^{-\i\tau L_1} U_m^+,
$$
where
\begin{equation*}
L_j=L_{\cal S}+ L_{\cal R}+ L_{\cal E}+V(\lambda)
\label{--1}
\end{equation*}
acts nontrivially on the subspace ${\cal H}_{\cal S} \otimes {\cal H}_{\cal R}\otimes {\cal H}_{{\cal E}_j}$ and
\begin{eqnarray*}
U_m^-&=&\exp\Big(-\i\tau \sum_{j=1}^m\sum_{k=1}^{j-1} L_{{\cal E}_k}\Big),\\
U_m^+&=&\exp\Big(-\i\tau \sum_{j=1}^m\sum_{k>j}^{j-1} L_{{\cal E}_k}\Big).
\end{eqnarray*}
Let $O_m$ be an instantaneous observable \fer{-7}. A straightforward computation shows that (see also \cite{bjm3}, equation (2.19))
\begin{equation}
\alpha^m(O_m) =(U_m^+)^* \e^{\i\tau L_1}\cdots\e^{\i\tau L_m} {\cal N}(O_m)            \e^{-\i\tau L_m}\cdots\e^{-\i\tau L_1} U_m^+,
\label{-40}
\end{equation}
with
\begin{equation}
{\cal N}(O_m) = A_{\cs\cR}\otimes_{j=-l}^{-1} \vartheta_{m+j}(\e^{\i \tau |j| L_\ce}B_j\e^{-\i \tau |j|L_\ce})\otimes_{j=0}^r\vartheta_{m+j}(B_j).
\label{-42}
\end{equation}

As normal states are convex combinations of vector states, it sufficient to consider the latters.
Let $\Psi_\rho$ be the GNS vector representing the initial state $\rho$, i.e., $\rho(\cdot) = \langle\Psi_\rho|\ \cdot\ \Psi_\rho\rangle$. It follows from the separating property of $\Psi_0$ (see \fer{-23}) that given any $\epsilon>0$, there is a $\widetilde B'\in{\frak M}'$ s.t. $\|\Psi_\rho -\widetilde B'\Psi_0\|<\epsilon$. Next, we approximate $\widetilde B'=B' +b_\epsilon(N)$, where $b_\epsilon(N)\rightarrow 0$ as $N\rightarrow\infty$ (for each $\epsilon$ fixed), and where $B'\in{\frak M}'$ has the form
\be
B'=B'_{\cal S}\otimes B'_{\cal R}\otimes_{n=1}^N B_n'\otimes_{n\geq N+1}\bbbone_{{\cal E}_n},
\label{mm9'}
\ee
with $B'_\#\in\mm'_\#$. The vector $B'\Psi_0$ is thus approximating the initial state $\Psi_\rho$. Let $O_m$ be an instantaneous observable and let us consider the expression
\begin{equation*}
\bra B'\Psi_0|\alpha^m(O_m) B'\Psi_0\ket=\bra\Psi_0|(B')^*B'\alpha^m(O_m) \Psi_0\ket.
\label{-mm1-}
\end{equation*}
We use expression \fer{-40} and the properties of the generators $K_n$ to obtain
\begin{equation*}
\bra\Psi_0|(B')^*B'\alpha^m(O_m) \Psi_0\ket = \bra\Psi_0|(B')^*B'(U^+_m)^* \e^{\i \tau K_1}\cdots \e^{\i\tau K_m}{\cal N}(O_m)\Psi_0\ket.
\label{-43}
\end{equation*}
Note that $U_m^+\Psi_0=\Psi_0$. Let
$$
P_N=\bbbone_\cS\otimes\bbbone_\cR\otimes \bbbone_{\cE_1}\otimes \cdots
\bbbone_{\cE_N}\otimes P_{\Psi_{\cE_{N+1}}}\otimes
P_{\Psi_{\cE_{N+2}}} \otimes \cdots,
$$
where $P_{\Psi_{\cE_k}}=|\Psi_{\cE_k}\ket\bra \Psi_{\cE_k} |$.
Since $(B')^*B'$ acts non-trivially only on the factors of the
chain Hilbert space having index $\leq N$, we have
\begin{eqnarray}
\lefteqn{
\bra\Psi_0|(B')^*B'\alpha^m(O_m) \Psi_0\ket=}\label{tri}\\
&&
\bra\Psi_0|(B')^*B'(\widetilde U_N^+)^*\e^{\i\tau K_1}\cdots \e^{\i\tau K_N} P_N\e^{\i\tau K_{N+1}}\cdots\e^{\i\tau K_m}{\cal N}(O_m) \Psi_0\ket,
\nonumber
\end{eqnarray}
where (for $m>N$; we have the limit $m\rightarrow\infty$ in mind)
$$
\widetilde U_N^+=P_NU_m^+ = \exp\Big( -\ri\tau \sum_{j=1}^{N-1} \sum_{k=j+1}^N  L_{\cE_k}\Big).
$$

Recall
\begin{equation*}
P_{\cs\cR}=\bbbone_{\cs}\otimes\bbbone_\cR\otimes|\Psi_\cc\ket\bra\Psi_\cc|.
\label{defp}
\end{equation*}
We have for $m>N+l$
\begin{eqnarray}
\!\!P_N\e^{\i\tau K_{N+1}}\cdots\e^{\i\tau K_m}{\cal N}(O_m) \Psi_0 &=& P_{\cs\cR}\e^{\i\tau K_{N+1}}\cdots\e^{\i\tau K_m}{\cal N}(O_m) \Psi_0\nonumber\\
&=& P_{\cs\cR} M^{m-l-N-1} \e^{\i\tau K_{m-l}}\cdots\e^{\i\tau K_m}{\cal N}(O_m)\Psi_0,\qquad
\label{-44}
\end{eqnarray}
where we have introduced the following {\it reduced dynamics
operator} (RDO), see (2.21) in \cite{bjm3}
\begin{equation}
P_{\cs\cR} \e^{\i\tau K}P_{\cs\cR} = M\otimes |\Psi_{\cal C}\ket\bra \Psi_{\cal C}|\simeq M \ \ \mbox{acting on} \ \
{\cal H}_{\cal S}\otimes {\cal H}_{\cal R}.
\label{-45}
\end{equation}
In the last step of \fer{-44}, we use the property
$$
P_{\cs\cR}\e^{\i\tau K_s}\e^{\i\tau K_{s+1}}\cdots\e^{\i\tau K_t}P_{\cs\cR} = P_{\cs\cR}\e^{\i\tau K_s}P_{\cs\cR}\e^{\i\tau K_{s+1}}P_{\cs\cR}\cdots P_{\cs\cR}\e^{\i\tau K_t}P_{\cs\cR},
$$
which holds for any $1\leq s< t$. This property follows from the independence of the $\cE_j$ for different $j$, see \cite{bjm1}, Proposition 4.1.

Combining \fer{tri} with \fer{-44} we obtain
\begin{eqnarray}
\lefteqn{
\bra\Psi_0|(B')^*B'\alpha^m(O_m) \Psi_0\ket=}\label{tri'}\\
&&
\bra\Psi_0|(B')^*B'(\widetilde U_N^+)^*\e^{\i\tau K_1}\cdots \e^{\i\tau K_N} P_{\cs\cR} M^{m-l-N-1} \e^{\i\tau K_{m-l}}\cdots\e^{\i\tau K_m}{\cal N}(O_m) \Psi_0\ket.
\nonumber
\end{eqnarray}

In order to emphasize the dependence on the coupling constants $\lambda=(\lambda_1,\lambda_2)$ we write $K(\lambda)$ and $M(\lambda)$. The following are general properties of the RDO.
\begin{prop}
Let $\lambda\in\R^2$ be arbitrary. We have

{\em i)} \ $M(\lambda)\in {\cal B}({\cal H}_{\cal S \cal R})$

{\em ii)} $M(\lambda) \Psi_{\cal SR}= \Psi_{\cal SR}$, where $\Psi_{\cal SR}:=\Psi_\cS\otimes\Psi_\cR$

\em{ iii)}  For any $\ffi$ in the dense set $\cD=\{A_{\cal SR}\Psi_{\cal SR}, \ A_{\cal SR}\in {\mathfrak M}_{\cal SR} \}$, there exists a constant $C(\ffi)<\infty$ s.t.
\be\label{alpow}
\sup_{n\in\N}\|M(\lambda)^n\ffi\|\leq  C(\ffi).
\ee
\end{prop}

\noindent
{\bf Proof:} i) follows from the fact that $K$ is a bounded perturbation of a self-adjoint operator and  ii) is a consequence of (\ref{kz}). To prove iii), first note that $\cD$ is dense since $\Psi_{\cs\cR}$ is cyclic for $\mm_{\cs\cR}$. Then note that the following identity holds for all $B_{\cs\cR}\in\mm_{\cs\cR}$
$$
\bra B_{\cal SR}\Psi_0|\alpha^n(A_{\cS\cR})\Psi_0\ket=\bra B_{\cal SR}\Psi_{\cal SR}|M(\lambda)^n A_{\cal SR}\Psi_{\cal SR}\ket.
$$
Statement iii) of the lemma follows from the density of $\cD$ and unitarity of the Heisenberg evolution, with $C(A_{\cal SR}\Psi_{\cal SR})=\|A_{\cal SR}\|$. \ep\\

\noindent
{\bf Remark.\ } Contrarily to the cases dealt with in \cite{bjm1}, \cite{bjm2} and \cite{bjm3}, where the underlying Hilbert space is finite dimensional, we cannot conclude from (\ref{alpow}) that $M(\lambda)$ is power bounded. Hence we do not know {\it a priori} that $\sigma(M(\lambda))\subset \{z\, : \, |z|\leq 1\}$.


\subsection{Translation analyticity}
\label{ssec:transanalytic}

To separate the eigenvalues from the continuous spectrum, we use analytic spectral deformation theory acting on the (radial) variable $s$ of the reservoir $\cR$.

Recall the definition \fer{mm9} of the translation. It is not difficult to see that
\be
K_\theta := T(\theta)^{-1}KT(\theta)= L_0+\theta N +\lambda_1W_{\cS\cR}(\theta)+\lambda_2 W_{\cS\cE},
\label{mm5}
\ee
where $N$ is the number operator,
and that the right side of \fer{mm5} admits an analytic continuation into $\theta\in\kappa_{\theta_0}$, strongly on the dense domain $D(L_0)\cap D(N)$, defining a family of closed operators (see \cite{JP}).

\begin{thm}[Analyticity of propagator]
\label{mmthm1}
Assume that H1 and H2 hold. Then
\medskip

1. $T(\theta)^{-1} \e^{\ri \tau K} T(\theta)$ has an analytic continuation from $\theta\in{\mathbb R}$ into the upper strip $\kappa_{\theta_0}$, and this continuation is strongly continuous as $\Im\theta\downarrow 0$.
\smallskip

2. For each $\theta\in\kappa_{\theta_0}\cup{\mathbb R}$, the analytic continuation of $T(\theta)^{-1} \e^{\ri \tau K} T(\theta)$ is given by $\e^{\ri\tau K_\theta}$, which is understood as an operator-norm convergent Dyson series (with `free part' $\e^{\ri\tau(L_0+\theta N)}$).
\smallskip

3. For each $\theta\in\kappa_{\theta_0}$, $\lambda_j\mapsto T(\theta)^{-1} \e^{\ri \tau K} T(\theta)$, $j=1,2$, are analytic entire functions.

\end{thm}

\noindent
{\bf Remarks.\ } 1. The proof of this result yields the following bound for all $\theta\in\kappa_{\theta_0}\cup\rx$,
$$
\|T(\theta)^{-1} \e^{\ri\tau K}T(\theta)\|\leq \e^{\tau \sup_{0\leq\Im\theta<\theta_0} \|W_\theta\|}.
$$

2. If $\theta_1, \theta_2\in\kappa_{\theta_0}$ with $\theta_1+\theta_2\in\kappa_{\theta_0}$, then
$$
T(\theta_1+\theta_2)^{-1} \e^{\ri\tau K}T(\theta_1+\theta_2) = T(\theta_2)^{-1}T(\theta_1)^{-1} \e^{\ri\tau K}T(\theta_1)T(\theta_2).
$$
In particular, $T(\theta)^{-1} \e^{\ri\tau K}T(\theta)$ is unitarily equivalent to $T(\ri\Im\theta)^{-1}\e^{\ri\tau K}T(\ri\Im\theta)$, via the unitary $T(\Re\theta)^{-1}$.

\bigskip

\noindent
{\bf Proof of Theorem \ref{mmthm1}.\ } For $\theta\in\rx$, the Dyson series expansion of $T(\theta)^{-1} \e^{\ri\tau K} T(\theta)$ is given by
\begin{eqnarray}
\lefteqn{
T(\theta)^{-1} \e^{\ri\tau K} T(\theta)}\label{mm2}\\
&&=\sum_{n=0}^\infty \ri^n\int_0^\tau\d t_1\cdots\int_0^{t_{n-1}}\d t_n \ \e^{\ri t_n\theta N}W_\theta(t_n) \e^{\ri(t_{n-1}-t_n)\theta N} W_\theta(t_{n-1}) \e^{\ri(t_{n-2}-t_{n-1})\theta N}\cdots\nonumber\\
&& \qquad\cdots \e^{\ri(t_1-t_2)\theta N} W_\theta(t_1)\e^{\ri(\tau-t_1)\theta N}\e^{\ri\tau L_0},\nonumber
\end{eqnarray}
where we define
$$
W_\theta(t)=\e^{\ri t L_0}W_\theta\e^{-\ri t L_0}
$$
In the derivation of \fer{mm2}, we use that for all $t\in\rx$,
$$
T(\theta)^{-1} \e^{\ri t L_0} T(\theta) = \e^{\ri t (L_0+\theta N)}=\e^{\ri t L_0}\e^{\ri t\theta N}=\e^{\ri t\theta N}\e^{\ri t L_0}.
$$
All the operators $\e^{\ri(t_{k-1}-t_k)\theta N}$, as well as $\e^{\ri t_n\theta N}$ and $\e^{\ri(\tau-t_1)\theta N}$ appearing in the integrand of \fer{mm2} have analytic extensions from real $\theta$ to $\theta\in\kappa_{\theta_0}$ which are continuous at $\rx$, and each of those extensions is bounded, having, in fact, norm one (uniformly in $\theta\in\kappa_{\theta_0}\cup \rx$ and in the $t_j$). Consequently, due to assumption H2, for fixed values of $t_1,\ldots,t_n$, the integrand in \fer{mm2} has an analytic extension into $\kappa_{\theta_0}$ which is again continuous on $\rx$. Let's call this extension $h_{t_1,\ldots,t_n}(\theta)$. For $\theta\in\kappa_{\theta_0}$, we have
\be
\|\partial_\theta h_{t_1,\ldots,t_n}(\theta)\|\leq C(\|\partial_\theta W_\theta\|+[\Im\theta]^{-1}),
\label{mm4}
\ee
uniformly in $t_1,\ldots,t_n$, for some constant $C$ (which depends on $n$). It thus follows, using the Lebesgue Dominated Convergence Theorem, that the integral on the r.h.s. of \fer{mm2} is analytic in $\theta\in\kappa_{\theta_0}$. To show \fer{mm4} we note that the derivative $\partial_\theta h$ is a sum of terms where $\partial_\theta$ is either applied to one of the $W_\theta$ or to an exponential. In the latter case, we have
$$
\|\partial_\theta \e^{\ri t\theta N}\| = \|tN\e^{-t\Im\theta N}\|=[\Im\theta]^{-1}\|t\Im\theta N\e^{-t\Im\theta N}\|\leq [\Im\theta]^{-1}\sup_{x\geq 0} x\e^{-x},
$$
where $\partial_\theta \e^{\ri t\theta N}$ is understood in the strong sense.

The norm of the integral on the r.h.s. of \fer{mm2} is bounded above by
$$
\frac{[\tau \sup_{0\leq\Im\theta<\theta_0} \|W_\theta\|]^n}{n!},
$$
uniformly in $\theta\in\kappa_\theta$. It follows that the series \fer{mm2} converges uniformly (Weierstrass $M$-test) and therefore the r.h.s. of \fer{mm2} is an analytic function in $\theta\in\kappa_{\theta_0}$.

Using the Lebesgue Dominated Convergence Theorem and the Weierstrass $M$-test as above, one readily shows that the series in \fer{mm2} is strongly continuous as $\Im\theta\downarrow 0$. Since equality \fer{mm2} holds for real $\theta$ this means that indeed the Dyson series is an analytic extension of $T(\theta)^{-1} \e^{\ri \tau K}T(\theta)$ into $\theta\in\kappa_{\theta_0}$. Note that the series is indeed the Dyson series of $\e^{\ri\tau K_\theta}$.

Finally, analyticity in $\lambda_1$ and $\lambda_2$ is clear from \fer{mm2} and \fer{mm20}. \hfill$\blacksquare$

\bigskip
The following result follows immediately from Theorem \ref{mmthm1} and definition \fer{-45}.
\begin{cor}
\label{mmcor1}
Recall the definition \fer{-45} of $M(\lambda)$. If assumption H2 is satisfied, then $T(\theta)^{-1} M(\lambda)T(\theta)$ has an analytic continuation into $\theta\in\kappa_{\theta_0}$, denoted $M_\theta(\lambda)$, and this continuation is continuous at $\rx$. We have $M_\theta(\lambda)\Psi_{\cS\cR}=\Psi_{\cS\cR}$.
\end{cor}


\subsection{Convergence to asymptotic state}

In order to make a link with the dynamics of observables, we
insert $\bbbone=T(\theta)T(\theta)^{-1}$ (with $\theta\in\rx$)
into equation \fer{tri'} to obtain
\begin{eqnarray}
\lefteqn{
\bra\Psi_0|(B')^*B'\alpha^m(O_m) \Psi_0\ket=}\label{mm30}\\
&&
\!\!\!\!\!\!\!\bra\Psi_0|(B')^*B'(\widetilde U_N^+)^*\e^{\i\tau K_1}\cdots \e^{\i\tau K_N} P_{\cs\cR} T(\theta) M_\theta^{m-l-N-1} T(\theta)^{-1}\e^{\i\tau K_{m-l}}\cdots\e^{\i\tau K_m}{\cal N}(O_m) \Psi_0\ket.
\nonumber
\end{eqnarray}
If $O_m$ is analytic (see definition \fer{-22}) then so is ${\cal N}(O_m)$ (see \fer{-42}), and therefore, by Theorem \ref{mmthm1}, $\e^{\i\tau K_{m-l}}\cdots\e^{\i\tau K_m}{\cal N}(O_m)$ is analytic as well (and continuous on the real axis). The r.h.s. of \fer{mm30} is thus an analytic function in $\theta\in\kappa_{\theta_0}$ (c.f. Theorem \ref{mmthm1} and Corollary \ref{mmcor1}). Moreover, this function is continuous on $\mathbb R$, and, by unitarity of $T(\theta)$ for real $\theta$, it is constant for real $\theta$. Hence \fer{mm30} is valid for all $\theta\in\kappa_{\theta_0}\cup\mathbb R$.

One expects that the operator $M_\theta(\lambda)^{m-l-N-1}$ converges to the projection onto the manifold of its fixed-points, as $m\rightarrow\infty$. Under certain (physically reasonable) conditions, this projection has rank one and is given by $|\Psi_{\cs\cR}\rangle\langle \psi^*_\theta(\lambda)|$, where $\psi^*_\theta(\lambda)$ is the unique invariant vector of the adjoint operator $[M_\theta(\lambda)]^*$, normalized as $\langle\Psi_{\cS\cR}|\psi^*_\theta(\lambda)\rangle =1$.

\begin{lem}
\label{limm} Assume that Condition FGR holds (see before
\fer{mm13}). Then for $0<|\lambda|<\lambda_0$,
$$
\lim_{n\ra
\infty}M_{\theta_1}(\lambda)^n=P_{1,M_{\theta_1}(\lambda)}
=|\Psi_{\cS\cR}\ket\bra\psi^*_{\theta_1}(\lambda)|,
$$
where $P_{1,M}$ denotes the spectral projector of the operator $M$ corresponding to the eigenvalue 1. The convergence is in operator norm, and occurs exponentially quickly.
\end{lem}

\noindent {\bf Proof.\ } Using FGR we have
$$
M=P_{1,M}+M_Q, \ \ \mbox{where} \ \ M_Q=QMQ, \ \ Q =\un -P_{1,M},
$$
with
$$
P_{1,M}=|\Psi_{\cS\cR}\ket\bra\psi^*|, \ \ \bra \psi^*|\Psi_{\cS\cR}\ket=1.
$$
Moreover, there exists $\gamma=\;\gamma_{\theta_1}(\lambda)>0$
s.t. \be \sigma(M_Q)\subset \{z\in\C \, : \,  |z| < e^{-\gamma}\}.
\label{mm21} \ee Therefore, the spectral radius of $M_Q$ satisfies
$$
\mbox{spr}\,(M_Q)=\lim_{n\ra\infty} \|M_Q^n\|^{1/n} <e^{-\gamma}<1.
$$
This, together with the identity
$$
M^n=P_{1,M}+M_Q^n
$$
yields for $n$ large enough,
$$
\|M^n-P_{1,M}\|\leq e^{-n\gamma}\ra 0 \  \ \mbox{as} \ \ n\ra \infty.
$$
(For some slightly smaller $\gamma$ than in \fer{mm21} above.)
\ep

\bigskip
It is now apparent from \fer{mm30} and Lemma \ref{limm} how to complete the proof of Theorem \ref{1st}: the increasing power of $M_\theta$ drives the system to an asymptotic state. Some care has to be exercised in the implementation of the complex deformation in the remaining part of the proof. Here are the details.

The approximation of the initial state $\rho$ by the vector $B'\Psi_0$, explained  in \fer{mm9'}, gives us the estimate
\begin{equation*}
\rho(\alpha^m(O_m)) = \bra\Psi_0|(B')^*B'(U^+_m)^* \e^{\i \tau L_1}\cdots \e^{\i\tau L_m}{\cal N}(O_m)\e^{-\i\tau L_m}\cdots\e^{-\i\tau L_1}\Psi_0\ket +R_1(O_m),
\end{equation*}
where $R_1$ is a bounded linear functional on $\frak M$ with $\lim_{N\rightarrow\infty}\|R_1\|<\epsilon$ (uniformly in $m$; note also that $\|O_m\|$ is independent of $m$). Here, $\epsilon$ is the approximation parameter fixed after \fer{-42}.
Let $\sigma\geq 0$ and define the spectral cutoff operator $\chi_\sigma:=\chi(|\d\Gamma(\partial_s)|\leq\sigma)$, acting (non-trivially only) on ${\cal H}_{\cR}$. (Here, $\chi(|x|\leq\sigma)$ equals one if $|x|\leq\sigma$ and zero otherwise.) The role of $\chi_\sigma$ is to smoothen the deformation operators: indeed, $\chi_\sigma T(\theta)^{-1}$ is analytic entire in $\theta$, see also \fer{mm9}. As $\sigma\rightarrow \infty$, $\chi_\sigma$ approaches the identity in the strong operator topology. Consequently, we have (see also \fer{tri'})
\begin{eqnarray}
\lefteqn{\rho(\alpha^m(O_m))}\nonumber\\
 &=&
 \langle\Psi_0|(B')^*B'(\tilde U_N^+)^*\chi_\sigma\e^{\ri\tau L_1}\cdots \e^{\ri\tau L_m} {\cal N}(O_m) \e^{-\ri\tau L_m}\cdots \e^{-\ri\tau L_1}\Psi_0\rangle +R_2(O_m)\nonumber\\
&=&\langle\Psi_0|(B')^*B'(\tilde U_N^+)^*\chi_\sigma\e^{\ri\tau K_1}\cdots \e^{\ri\tau K_N} M^{m-l-N-1} \e^{\i\tau K_{m-l}}\cdots\e^{\i\tau K_m}{\cal N}(O_m) \Psi_0\rangle \nonumber\\
&& +R_2(O_m),
\label{mm10'}
\end{eqnarray}
where $R_2$ is a bounded linear functional on $\frak M$ satisfying $\lim_{\sigma\rightarrow \infty}R_2=R_1$, and in particular, $\lim_{N\rightarrow\infty}\lim_{\sigma\rightarrow\infty}\|R_2\|<\epsilon$ (uniformly in $m$).

We now introduce the spectral deformation in the main term on the r.h.s. of \fer{mm10'}. We have for all $\theta\in\mathbb R$
\begin{eqnarray}
\lefteqn{
\langle\Psi_0|(B')^*B'(\tilde U_N^+)^*\chi_\sigma\e^{\ri\tau K_1}\cdots \e^{\ri\tau K_N}  M^{m-l-N-1} \e^{\i\tau K_{m-l}}\cdots\e^{\i\tau K_m}{\cal N}(O_m) \Psi_0\rangle}\nonumber\\
&&\qquad = \langle\Psi_0|(B')^*B'(\tilde U_N^+)^*\chi_\sigma T(\theta) \e^{\ri\tau K_{1,\theta}}\cdots \e^{\ri\tau K_{N,\theta}}  M_\theta^{m-l-N-1}\times\nonumber\\
&&\qquad\ \ \ \ \times  T(\theta)^{-1} \e^{\i\tau K_{m-l}}\cdots\e^{\i\tau K_m}{\cal N}(O_m)\Psi_0\rangle,
\label{mm10}
\end{eqnarray}
where $K_{j,\theta}=T(\theta)^{-1} K_j T(\theta)$. Since $O_m$ is
an analytic observable, and according to Theorem \ref{mmthm1}, the
right side of \fer{mm10} has an analytic continuation into
$\theta\in\kappa_{\theta_0}$ and this continuation is continuous
at $\mathbb R$. Moreover, on $\mathbb R$, this continuation is a
{\it constant} function (equal to the left side of \fer{mm10}). It
follows that the analytic continuation is constant on the whole
region of analyticity plus the real axis, and \fer{mm10} holds for
all $\theta\in\kappa_{\theta_0}\cup\mathbb R$. Due to the
Condition FGR and Lemma \ref{limm}, we have
\begin{eqnarray}
\lefteqn{
\langle\Psi_0|(B')^*B'(\tilde U_N^+)^*\chi_\sigma\e^{\ri\tau K_1}\cdots \e^{\ri\tau K_N} M^{m-l-N-1} \e^{\i\tau K_{m-l}}\cdots\e^{\i\tau K_m}{\cal N}(O_m) \Psi_0\rangle}\nonumber\\
&=& \langle\Psi_0|(B')^*B'(\tilde U_N^+)^*\chi_\sigma T(\theta_1) \e^{\ri\tau K_{1,\theta_1}}\cdots \e^{\ri\tau K_{N,\theta_1}} \Psi_0\rangle \times\nonumber\\
&&\ \ \ \ \times \langle\psi_{\theta_1}^*| T(\theta_1)^{-1} \e^{\i\tau K_{m-l}}\cdots\e^{\i\tau K_m}{\cal N}(O_m) \Psi_0\rangle +\e^{-m\gamma} R_3(O_m)\nonumber\\
&=& \|B'\Psi_0\|^2 \langle \psi_{\theta_1}^*| T(\theta_1)^{-1} \e^{\i\tau K_{m-l}}\cdots\e^{\i\tau K_m}{\cal N}(O_m)\Psi_0\rangle +\e^{-m\gamma} R_3(O_m),
\label{mm11}
\end{eqnarray}
where $\gamma>0$ (see proof of Lemma \ref{limm}) and where
$$
\|R_3(O_m)\|\leq C(\theta_0,N)\|T(\theta_1)^{-1} \e^{\i\tau K_{m-l}}\cdots\e^{\i\tau K_m}{\cal N}(O_m)\Psi_0\|\ \e^{\theta_1\sigma} \e^{N\gamma}.
$$
The latter quantity is bounded uniformly in $m$.
To arrive at the second line in \fer{mm11} we made use of the fact that $\Psi_0$ is invariant under the action of all of $\chi_\sigma$, $T(\theta)$ and $\e^{\i\tau K_{j,\theta}}$.
We combine estimates \fer{mm10'},  \fer{mm10} and \fer{mm11} to arrive at
\begin{eqnarray}
\lefteqn{
\limsup_{m\rightarrow\infty}\left|\rho(\alpha^m(O_m)) - \langle \psi_{\theta_1}^*| T(\theta_1)^{-1} \e^{\i\tau K_{m-l}}\cdots\e^{\i\tau K_m}{\cal N}(O_m)\Psi_0\rangle \right|}\nonumber\\
&& \leq \big|\|B'\Psi_0\|^2 -1\big|\ \sup_m | \langle \psi_{\theta_1}^*| T(\theta_1)^{-1} \e^{\i\tau K_{m-l}}\cdots\e^{\i\tau K_m}{\cal N}(O_m)\Psi_0\rangle| \nonumber\\
&& \qquad +\limsup_{m\rightarrow\infty}|R_2(O_m)|.
\label{mm12}
\end{eqnarray}
By taking in \fer{mm12} first $\sigma\rightarrow\infty$ and then $N\rightarrow\infty$, we see that the r.h.s. of \fer{mm12} is bounded above by $\epsilon$ (note that in this double limit, $\|B'\Psi_0\|$ tends to $\|\Psi_\rho\|=1$). Since $\epsilon$ was chosen arbitrarily small to begin with, it follows that the left hand side of \fer{mm12} is zero, and thus \fer{mm13} follows. This completes the proof of Theorem \ref{1st}.\hfill$\blacksquare$

\section{Analysis of $M(\lambda)$}
\label{sec:rdoanalysis}

An important issue in the analysis of concrete models is the
verification of the {\it Fermi Golden Rule} assumption FGR (see
before Theorem \ref{1st}). We have introduced the description of
the two types of systems, `small' and `reservoir' in Section
\ref{sec:descripintro}. For a more detailed analysis, we need to
complement that description.

We denote eigenvalues and associated eigenvectors of $H_\cs$ by
$$
E_1, \cdots, E_d, \ \ \ \mbox{and } \ \ \ \ffi_1, \cdots, \ffi_d.
$$
Before analyzing the spectrum of $M_\theta(\lambda)$ in general, we mention some easier special cases.

$\bullet$ In the unperturbed case ($\lambda=0$) we have
\begin{eqnarray*}
M(0)&=&\e^{\i\tau L_{\cal S}}\otimes \e^{\i\tau L_{\cal R}}, \ \ \mbox{with}\\
\sigma(M(0))&=&\left\{\e^{\i\tau(E_j-E_k)}\right\}_{ (j,k)\in
\{1,2,\cdots, d\}^2}\cup \left\{\e^{\i l}, \  l\in\R\right\},
\end{eqnarray*}
where the eigenvalues  $\e^{\i\tau(E_j-E_k)}$ are embedded and have corresponding eigenvectors $\ffi_j\otimes \ffi_k\otimes \Psi_{\cal R}$. The eigenvalue 1 is at least $d$-fold degenerate.

$\bullet$ In case the coupling  $\lambda_1$ between the small system and the reservoir is zero, we have
$$
M(0,\lambda_2)=\tilde M(\lambda_2)\otimes \e^{\i\tau L_R} \ \ \mbox{on} \ \ \cH_\cS\otimes\cH_{\cR},
$$
where
$$
\tilde M(\lambda_2)\simeq P_\cS \e^{\i\tau(L_\cS+L_\cE+\lambda_2V_{\cal SE} )}P_\cS \ \ \mbox{and} \ \ P_\cs=\I_\cS\otimes|\Psi_\cE\ket\bra\Psi_\cE |.
$$
The results of \cite{bjm1} apply to $\tilde M(\lambda_2)$, which
is nothing but the RDO corresponding to the repeated interaction
quantum system formed by $\cS$ and $\cC$ only. In particular, we
get that $M(0,\lambda_2)$ is power bounded, as $\tilde
M(\lambda_2)$ is and $e^{i\tau L_R}$ is unitary. Moreover,
assuming the interaction $V_{\cal SE}$ ``effectively'' couples
$\cS$ and $\cC$, hypothesis (E) in \cite{bjm1},  we know that the
spectrum of $\tilde M(\lambda_2)$ satisfies
$$ \sigma(\tilde
M(\lambda_2))=\{\mu_j(\lambda_2)\}_{ \ j=1,2,\cdots, d^2},
$$
with $\mu_1(\lambda_2)=1$ a simple eigenvalue with eigenvector
$\Psi_\cS$ and $\mu_j(\lambda_2)\in  \{z \, | \, |z|< 1\}$. Hence,
$$
\sigma(M(0,\lambda_2))=\sigma(\tilde M(\lambda_2))\cup
\{|\mu_j(\lambda_2)|\e^{\i l}, \ l\in\R\}_{j=1,\cdots, d^2},
$$
where the eigenvalues are embedded in the absolutely continuous
spectrum again.

$\bullet$ In case the chain is decoupled, i.e. if $\lambda_2=0$, we get
$$
M(\lambda_1,0)= \e^{\i\tau (L_\cS+L_\cR+\lambda_1W_{\cS\cR})}   \ \ \mbox{on} \ \ \cH_\cS\otimes\cH_{\cR},
$$
whose spectral analysis already requires the tools we will use for general $\lambda$.

\bigskip
We now turn to a perturbative analysis of $M_\theta(\lambda)$ (small $\lambda$).
Take $\theta\in\kappa_{\theta_0}$ and let $\lambda_1=\lambda_2=0$. Then
$$
M_{\theta}(0)=\e^{\ri\tau(L_\cS+L_\cR+\theta N)}=\e^{\ri\tau L_\cS}\otimes \e^{\ri\tau L_\cR}\e^{\ri\tau\theta N}
$$
and
\begin{equation*}\label{smo}
\sigma({M_{\theta}(0)})=\{ \e^{\ri\tau(E_j-E_k)}\}_{j,k\in
\{1,\cdots, d\}}\cup\{\e^{\ri l}\e^{-\tau j\Im\theta}, \ l\in\R
\}_{j\in \N^*}.
\end{equation*}
The effect of the analytic translation is to push the continuous
spectrum  of $M_{\theta}(0)$ onto circles with radii $\e^{-\tau
j\Im\theta}$, $j=1,2,\ldots$, centered at the origin. Hence the
discrete spectrum of $M_\theta(0)$, lying on the unit circle, is
separated from the continuous spectrum by a distance
$1-e^{-\tau\Im \theta}$. Analytic perturbation theory in the
parameters $\lambda_1, \lambda_2$ guarantees that the discrete and
continuous spectra stay separated for small coupling. The
following result quantifies this.

\begin{prop}
\label{spec}
Let $C_0(\lambda):=\sup_{\theta\in\kappa_{\theta_0}\cup \mathbb R}\|W_\theta\|$. Take $\theta\in\kappa_{\theta_0}$ and suppose that
\be
\tau C_0(\lambda)\e^{\tau C_0(\lambda)} < \textstyle\frac 14 (1-\e^{-\tau\Im\theta}).
\label{mm25}
\ee
Then the spectrum of $M_\theta(\lambda)$ splits into two disjoint parts,
$$
\sigma(M_{\theta}(\lambda))=\sigma^{(0)}_{\theta}(\lambda)\cup
\sigma^{(1)}_{\theta}(\lambda) \ \ \mbox{with } \ \
\sigma^{(0)}_{\theta}(\lambda)\cap
\sigma^{(1)}_{\theta}(\lambda)=\emptyset.
$$
These parts are localized as follows
\bea
\sigma^{(0)}_{\theta}(\lambda)&\subset& \big\{z \ : \ 1-\textstyle\frac 14 (1-\e^{-\tau \Im\theta})< |z|\leq 1\big\} \label{mm27}\\
\sigma^{(1)}_{\theta}(\lambda)&\subset& \big\{z \ : \ 0\leq |z|< \e^{-\tau \Im\theta}+\frac14 (1-\e^{-\tau \Im\theta})
\big\}.\label{mm28}
\eea
Moreover, the spectrum $\sigma_\theta^{(0)}(\lambda)$ is purely discrete, consisting of $d^2$ eigenvalues (counted including algebraic multiplicities).
\end{prop}

\noindent
{\bf Proof.\ }
According to the Dyson series expansion \fer{mm2}, we have
\be
M_\theta(\lambda) = \e^{\ri\tau(L_0+\theta N)} + S_2,
\label{mm23}
\ee
where
$$
\|S_2\|\leq \sum_{n=1}^\infty \frac{[\tau C_0(\lambda)]^n}{n!} \leq \tau C_0(\lambda) \e^{\tau C_0(\lambda)}.
$$
Since $\e^{\ri\tau(L_0+\theta N)}$ is a normal operator, the spectrum of the perturbed operator $M_\theta(\lambda)$, \fer{mm23}, lies inside a set whose distance to $\sigma(\e^{\ri\tau(L_0+\theta N)})$ does not exceed $\|S_2\|$. The spectrum of $\e^{\ri\tau(L_0+\theta N)}$ consists of isolated eigenvalues lying on the unit circle and of continuous spectrum lying on concentric circles centered at the origin, with radii $\e^{-\tau n\Im\theta}$, $n=1,2,\ldots$

It follows that if \fer{mm25} is satisfied, then the continuous spectrum of $M_\theta(\lambda)$ is located as in \fer{mm28}. Furthermore, the spectral radius of $M_\theta(\lambda)$ is determined by discrete eigenvalues only, and so by Lemma \ref{spr} below, these eigenvalues cannot lie outside the unit circle, from which \fer{mm27} follows. \hfill $\blacksquare$

\medskip
\noindent {\bf Remark.\ } We always have
$1\in\sigma_\theta^{(0)}(\lambda)$, with eigenvector
$\Psi_{\cS\cR}$.

As a consequence of Proposition \ref{spec}, a verification of FGR
for concrete models, like the one of Section \ref{ssec:explexa},
is done via (perturbative) analysis \emph{only of the discrete
eigenvalues} of $M_\theta(\lambda)$.


\section{Energy fluxes, entropy production}
\label{sectvarenergy}

We use the notation and definitions of Section \ref{sectthermo} and assume thoughout that Assumption H3 is satisfied. We have
$$
\alpha^{m+1}(L_\cR)-\alpha^m(L_\cR)=\alpha^m(\alpha^\tau_{m+1}(L_\cR)-L_\cR),
$$
with
$$
\alpha^\tau_{m}(\cdot)=\e^{\i\tau\widetilde{L}_m}\,\cdot\, \e^{-\i\tau\widetilde{L}_m},\qquad \widetilde L_m = L_0+\vartheta_m(V).
$$
Thus
$$
\alpha^\tau_{m+1}(L_\cR) - L_\cR = \i\int_0^\tau \alpha^t_{m+1}([\widetilde{L}_{m+1}, L_\cR])\,\d t =\i\int_0^\tau \alpha^t_{m+1}([\lambda_1V_{\cs\cR}, L_\cR])\,\d t,
$$
and we arrive at the expression for the variation of energy in the reservoir
$$
\Delta E^{\cR}(m)=\alpha^m\left(\i\int_0^\tau
\alpha^t_{m+1}([\lambda_1 V_{\cs\cR}, L_\cR])\,\d t\right).
$$
{}From definition \fer{-7} and assumption {\bf H3} (see before
\fer{-55}) together with the analyticity of the dynamics (Theorem
\ref{mmthm1}), it is clear that $\alpha^t_{m+1}([\lambda_1
V_{\cs\cR}, L_\cR])$ is an analytic instantaneous observable. By
approximating the integral $\int_0^\tau \d t$ by a Riemann sum
(converging uniformly in $m$) we see that the integral in question
is the limit of a sum of instantaneous observables all having
uniformly bounded indices $l,r$ (see \fer{-7}). Furthermore,
$\alpha^m$ is bounded uniformly in $m$ and therefore we can apply
Theorem \ref{1st} to conclude that
\begin{equation}
\lim_{m\rightarrow\infty}\rho\left(\Delta E^{\cR}(m)\right) =
\rho_{+,\lambda}\left( \i \int_0^\tau \alpha_{m+1}^t(\lambda_1
[V_{\cs\cR}, L_\cR] \d t \right).
 \label{mgren1}
\end{equation}
Next we examine the variation of energy in the chain $\cC$. During
the time interval $[m\tau, (m+1)\tau)$, the energy of the element
$\cE_k$, $k\neq m+1$, of the chain is invariant. Hence, the
variation of energy in the whole chain between the times $m$ and
$m+1$ coincides with that of the element $\cE_{m+1}$ only. Thus,
$$
\Delta E^{\cC}(m)=\alpha^{m+1}(L_{\cE_{m+1}})-L_{\cE_{m+1}}\equiv \alpha^{m+1}(L_{\cE_{m+1}})-\alpha^m(L_{\cE_{m+1}}).
$$
Proceeding as above, we arrive at
$$
\Delta E^{\cC}(m)=\alpha^m\left(\i\int_0^\tau \alpha^t_{m+1}([\lambda_2 \vartheta_{m+1}(V_{\cs\cE}), L_{\cE_{m+1}}])\,\d t\right),
$$
where $\i\int_0^\tau \alpha^t_{m+1}([\vartheta_{m+1}(V_{\cs\cE}), L_{\cE_{m+1}}])\,\d t$ is an instantaneous observable (Assumption H3).

Finally, for the variation of energy in the small system $\cs$ we
get
$$
\Delta E^{\cs}(m)=\alpha^m\left(\i \int_0^\tau \alpha^t_{m+1}([\lambda_2\vartheta_{m+1}(V_{\cs\cE})+\lambda_1V_{\cs\cR}, L_\cS])\,\d t\right).
$$

It follows now from Theorem \ref{1st} (by the same reasoning
leading to \fer{mgren1}, see also \cite{bjm1})
\begin{equation*}\label{totalenergyprod}
\d E_+^{\mbox{\scriptsize
tot}}=\rho_{+,\lambda}(j^{\mbox{\scriptsize
tot}})=\rho_{+,\lambda}(V-\alpha^\tau(V)),
\end{equation*}
where \be j^{{\rm tot}}=V-\alpha^\tau(V)=-\i\int_0^\tau
\alpha^t([L_\cs+L_\cR+L_\ce,\lambda_2
V_{\cs\ce}+\lambda_1V_{\cs\cR}])\, \d t. \label{-59} \ee

On the other hand, by the same reasoning,
$\d E_+^{\#}$, for $\#\, = \cS, \cE, \cR$ (see \fer{-55}),  is given by $\rho_{+,\lambda}(j^\#)$, where
\bea
j^\cS&=&\i\int_0^\tau \alpha^t([\lambda_2V_{\cs\cE}+\lambda_1V_{\cs\cR}, L_\cS])\,\d t, \label{energyobs}\\
j^{\cE}&=&\i\int_0^\tau \alpha^t([\lambda_2V_{\cs\cE}, L_\cE])\,\d t,\label{enobs1}\\
j^{\cR}&=&\i\int_0^\tau \alpha^t([\lambda_1V_{\cs\cR}, L_\cR])\,\d
t. \label{enobs2} \eea

The following result relates the various flux observables $j^\#$.
\begin{prop}
\label{fluxprop}
We have $j^{\mbox{\scriptsize tot}}=j^\cS+j^{\cE}+j^{\cR}$. Furthermore, $\d E_+^\cs:=\rho_{+,\lambda}(j^\cs)=0$.
\end{prop}

\noindent
{\bf Proof.} The relation $j^{\mbox{\scriptsize tot}}=j^\cS+j^{\cE}+j^{\cR}$ follows directly from \fer{-59} and \fer{energyobs}-\fer{enobs2}. To see that $\d E_+^\cs=0$ we note that since $L_\cs $ is bounded we have
$$
\frac1N(\alpha^{N}(L_\cs)-L_\cs)=\frac1N\sum_{m=1}^N \alpha^{m}(L_\cs)- \alpha^{m-1}(L_\cs)\ra 0, \ \ \ \mbox{as} \ \ \ N\ra \infty.
$$
\ {} \hfill $\blacksquare$

\medskip
The main ingredient in the analysis of the entropy production is the following entropy production formula, established in \cite{jp2003}
\begin{eqnarray*}
\lefteqn{ \Delta S(m)}\nonumber\\
&& =  \rho\Big(U(m)^*\Big[\beta_\cE\sum_k L_{\cE,k} +\beta_\cS L_\cS+\beta_\cR L_\cR \Big]U(m)   -\beta_\cE\sum_k L_{\cE,k} -\beta_\cS L_\cS -\beta_\cR L_\cR\Big).
\end{eqnarray*}
Following the proof of Proposition 2.6 of \cite{bjm1}, we get
\bea
\Delta S(m)\!\! & = & \!\!(\beta_\cR-\beta_\cE) \sum_{k=1}^{m-1} \rho(\Delta E^\cR(k))+ (\beta_\cS-\beta_\cE) \sum_{k=1}^{m-1} \rho(\Delta E^\cS(k))+\beta_\cE \sum_{k=1}^m \rho(\Delta E^{{\rm tot}}(k))\nonumber\\
 & & + \beta_\cE \left[ \rho(\lambda_1V_{\cS\cR})-\rho(\alpha^m (\lambda_1V_{\cS\cR})) + \rho(\lambda_2\vartheta_1(V_{\cS\cE}))-\rho(\alpha^m(\lambda_2\vartheta_{m+1}(V_{\cS\cE}))) \right].\nonumber
\eea
Hence using Proposition \ref{fluxprop}, and since $\rho(\alpha^m (V_{\cS\cR}))$ and $\rho(\alpha^m(\vartheta_{m+1}(V_{\cS\cE})))$ are bounded in $m$, we get
\be\label{eq:entropyprod}
\d S_+:= \lim_{m\to\infty} \frac{\Delta S(m)}{m}= (\beta_\cR-\beta_\cE) \d E_+^\cR + \beta_\cE \d E^{{\rm tot}}_+.
\ee


\section{More detail on the concrete example}
We consider the model described in Section \ref{ssec:explexa}. In
order to write down explicitly all interaction operators appearing
in the C-Liouvillean, we need the explicit form of modular data of
$\cS$, $\cR$ and $\cE$.

The modular data of $\cS$ and the elements $\cE$ of the chain
associated to the  reference states $\rho_\cS$ and
$\rho_{\beta_\cE,\cE}$ are given by
\begin{equation*}\label{jdsyst}
J_\s (\phi\otimes\psi) =\bar{\psi}\otimes\bar{\phi}, \quad \Delta_\s=\one_{\C^2}\otimes\one_{\C^2}.
\end{equation*}
\begin{equation*}\label{jdchain}
J_\ce (\phi\otimes\psi) =\bar{\psi}\otimes\bar{\phi}, \quad \Delta_\ce=\e^{-\beta_\cE L_\ce}.
\end{equation*}
The one of $\cR$ is given as follows (see also Theorem 3.3 of \cite{jp2002})
\begin{eqnarray*}
\label{jd}
J_\cR \ffi(f_{\beta_\cR}) J_\cR &=&\i\Gamma(-\bbbone)\ffi(f_{\beta_\cR}^\#)\\
\Delta_\cR&=&e^{-\beta_\cR L_\cR}\\
J_R&=&(-1)^{N(N+1)/2}{\cal C}\circ {\cal F}.
\end{eqnarray*}
Here, we have introduced the notation $f_\beta^\#(s) = \i\e^{-\beta s/2}f_\beta(s)=\bar f_\beta(-s)$, where the bar indicates the complex conjugate. Furthermore, $N=\d\Gamma(\bbbone)$ is the number operator, ${\cal C}$ is the complex conjugation operator and ${\cal F}$ is the sign flip operator acting on $f\in \otimes_{j=1}^nL^2(\rx,{\frak G})$ as
$$
({\cal F}f) (s_1,s_2, \dots, s_n)= f (-s_1,-s_2, \dots, -s_n).
$$

An easy computation leads to the following expression for the ``interaction'' part $W$ of the C-Liouvillean,
\begin{eqnarray*}
W(\lambda) & = & \lambda_1 \Big( \sigma_x \otimes \one_{\C^2} \otimes \varphi(f_{\beta_\cR}(s)) \\
&& \qquad- \one_{\C^2}\otimes\sigma_x \otimes \Gamma(-\one)  \left(a^*(f_{\beta_\cR}(s))-a(\e^{-\beta_\cR s}f_{\beta_\cR}(s))\right) \Big)\\
 & & + \lambda_2 \Big(a_\cS\otimes\one_{\C^2}\otimes a^*_\cE \otimes \one_{\C^2}+a^*_\cS\otimes\one_{\C^2}\otimes a_\cE \otimes \one_{\C^2}\\
 & & \qquad\left.-\e^{\beta_\cE E_\cE/2}\one_{\C^2}\otimes a^*_\cS \otimes \one_{\C^2}\otimes a_\cE - \e^{-\beta_\cE E_\cE/2}\one_{\C^2}\otimes a_\cS \otimes \one_{\C^2}\otimes a^*_\cE \right).
\end{eqnarray*}



Assumption (H4) on the form factor $f$ ensures that assumption (H3) of Section \ref{ssec:transanalytic} is satisfied with $\theta_0=\delta$. We can thus apply the general results of Section \ref{sec:rdoanalysis}. In particular, the map $\theta\mapsto T(\theta)^{-1}M(\lambda)T(\theta)$ has an analytic continuation in the strip $\kappa_{\theta_0}$ (see Corollary \ref{mmcor1}). We then fix some $\theta_1\in \kappa_{\theta_0}$ such that $1-\e^{-\tau \Im(\theta_1)}>0$. For $\lambda$ small enough eq. (\ref{mm25}) is therefore satisfied, so that we can verify the (FGR) hypotheses using perturbation theory for a finite set of eigenvalues, those four eigenvalues which are located in $\sigma_\theta^{(0)}(\lambda)$ (see (\ref{mm27})).
When the coupling constants are turned off, we have
$$
\sigma_\theta^{(0)}(0)=\sigma(\e^{\i\tau L_\cS})=\{1,\e^{\i\tau E_\cS},\e^{-\i\tau E_\cS}\}
$$
where the eigenvalue $1$ has multiplicity $2$. In order to make
the computation in perturbation theory as simple as possible, we
will assume that these eigenvalue do not coincide, i.e. $\tau
E_\cS \notin \pi\N$. However, this assumption is certainly not
necessary.

Using a Dyson expansion for $M_\theta(\lambda)$ as in the proof of Theorem \ref{mmthm1} and regular perturbation theory (see e.g. \cite{HP,kato}) we compute the four elements of $\sigma_\theta^{(0)}(\lambda)$. We know that $1$ always belongs to $\sigma_\theta^{(0)}(\lambda)$. The other ones respectively write
\begin{eqnarray*}
e_0(\lambda) & = & 1-\lambda_1^2 \tau \frac{\pi}{2}\sqrt{E_\cS} \|f(\sqrt{E_\cS})\|_{\mathfrak G}^2-\lambda_2^2 \tau^2 \sinc^2\left(\frac{\tau(E_\cE-E_\cS)}{2}\right)+O(\lambda^3),\\
e_+(\lambda) & = & \e^{i\tau E_\cS}\left[1-\lambda_1^2 \tau\frac{\pi}{4}\sqrt{E_\cS} \|f(\sqrt{E_\cS})\|_{\mathfrak G}^2-\lambda_2^2 \frac{\tau^2}{2} \sinc^2\left(\frac{\tau(E_\cE-E_\cS)}{2}\right)  \right.\\
 & & \left. -i \left(\lambda_1^2\frac{\tau}{4} \, {\rm PV} \, \int_\R \frac{\sqrt{|s|}\|f(\sqrt{|s|})\|_{\mathfrak G}^2}{s-E_\cS}\d s + \lambda_2^2 \tau^2 \frac{1-\sinc(\tau(E_\cE-E_\cS))}{\tau(E_\cE-E_\cS)}\right) \right]+O(\lambda^3),\nonumber\\
e_-(\lambda) & = & \e^{-i\tau E_\cS}\left[1-\lambda_1^2 \tau \frac{\pi}{4}\sqrt{E_\cS}\|f(\sqrt{E_\cS})\|_{\mathfrak G}^2-\lambda_2^2 \frac{\tau^2}{2} \sinc^2\left(\frac{\tau(E_\cE-E_\cS)}{2}\right)  \right.\\
 & & \left. +i \left(\lambda_1^2\frac{\tau}{4} \, {\rm PV} \, \int_\R \frac{\sqrt{|s|}\|f(\sqrt{|s|})\|_{\mathfrak G}^2}{s-E_\cS}\d s + \lambda_2^2 \tau^2 \frac{1-\sinc(\tau(E_\cE-E_\cS))}{\tau(E_\cE-E_\cS)}\right) \right]+O(\lambda^3),\nonumber
\end{eqnarray*}
where $\sinc(x)=\frac{\sin(x)}{x}$ and ${\rm PV}$ stands for Cauchy's principal value. We thus get the following
\begin{lem} Assume that $\|f(\sqrt{E_\cS})\|_{\mathfrak G}\neq 0$ and $\tau(E_\cE-E_\cS)\notin 2\pi \Z^*$, then (FGR) is satisfied.
\end{lem}

In order to compute the asymptotic state $\rho_{+,\lambda}$, we compute the (unique) invariant vector $\psi_{\theta}^*(\lambda)$ of $M_\theta(\lambda)^*$ (see (\ref{mm13})). Once again, standard perturbation theory shows that $\psi_\theta^*(\lambda)=\psi_\cS^*(\lambda) \otimes \Psi_\cR+O_\theta(\lambda)$ with
\begin{eqnarray*}
\psi_\cS^*(\lambda) & = & \sqrt{2}\frac{\lambda_1^2\gamma_{\rm th}^{(2)}Z_{\beta_\cR,\cS}^{-1}+\lambda_2^2 \gamma_{\rm ri}^{(2)}Z_{\beta_\cE',\cS}^{-1}}{\lambda_1^2 \gamma_{\rm th}^{(2)}+\lambda_2^2 \gamma_{\rm ri}^{(2)}}\; |00\ket \\
 & & +\sqrt{2}\frac{\lambda_1^2 \gamma_{\rm th}^{(2)}\e^{-\beta_\cR E_\cS}Z_{\beta_\cR,\cS}^{-1}+\lambda_2^2 \gamma_{\rm ri}^{(2)}\e^{-\beta_\cE' E_\cS}Z_{\beta_\cE',\cS}^{-1}}{\lambda_1^2 \gamma_{\rm th}^{(2)}+\lambda_2^2 \gamma_{\rm ri}^{(2)}}\; |11\ket,
\end{eqnarray*}
where $\gamma_{\rm th}^{(2)}$ and $\gamma_{\rm ri}^{(2)}$ are
defined in \fer{thermrate}-\fer{rirate}. Inserting the above
expression in \fer{-9}, this proves Proposition
\ref{ssf:asymptstate}.


\appendix
\renewcommand{\theequation}{A.\arabic{equation}}
\renewcommand{\theresultcounter}{A.\arabic{resultcounter}}

\section{Some operator theory}

Our analysis of the spectrum of $M_\theta(\lambda)$ makes use of a
translated version of (\ref{alpow}), which replaces the
powerboundedness of $M_\theta(\lambda)$ in our setup.
\begin{lem}\label{quasibdd}
Assume $A_{\cal SR}$ and $B_{\cal SR}$ are translation analytic in $\kappa_{\theta_0}$. Then
$$
\sup_{m\in\N}|\bra B_{\cal SR}\Psi_{\cS\cR}|M_\theta^m(\lambda)A_{\cal SR}\Psi_0\ket|\leq \|A_{\cal SR}(\theta)\|
\|B_{\cal SR}(\overline{\theta})\|.
$$
\end{lem}
{\bf Proof:}
Consider
\begin{eqnarray*}
\bra B_{\cal SR}(\overline{\theta})\Psi_{\cS\cR}|\alpha^m(A_{\cal SR}(\theta))\Psi_{\cal SR}\ket
&=&\bra B_{\cal SR}\Psi_{\cS\cR}|T(\theta)M(\lambda)^mT^{-1}(\theta)A_{\cal SR}\psi_{\cal SR}\ket\nonumber\\
&=&\bra B_{\cal SR}\Psi_{\cS\cR}|M_\theta(\lambda)^mA_{\cal SR}\psi_{\cal SR}\ket
\end{eqnarray*}
\ep

We can use this property to bound the spectral radius of
$M_\theta(\lambda)$ when it is determined by discrete eigenvalues
only. This  means that there are finitely many eigenvalues
$\alpha_j$, $j=1,\ldots,N$, all of equal modulus $\alpha$, such
that $\sup\{ |z|\, : \, z\in\sigma(M_\theta(\lambda))\}=\alpha$
and $\sigma_{{\rm ess}}(M_\theta(\lambda)) \cap
\{|z|=\alpha\}=\emptyset$.

\begin{lem}\label{spr}
Assume that for some $\theta\in \kappa_{\theta_0}$,  $\mbox{\em spr}\,(M_{\theta}(\lambda))$ is determined by discrete eigenvalues only.
Then $\mbox{\em spr}\,(M_{\theta}(\lambda))= 1$.
\end{lem}

This is an application of the following result stated in a more abstract setting.

\begin{prop}
\label{draco}
Let $M$ be a bounded operator on a Hilbert space ${\cal H}$ such that:\\
i) there exists a dense set of vectors ${\cal C}\subset {\cal H}$ satisfying
$$\sup_{n\in\N}|\bra \ffi |M^n \psi \ket|\leq C(\ffi,\psi), \ \ \ \forall \ffi, \psi \in {\cal C}, $$
ii)  $\mbox{\em spr}\,(M)$ is determined by discrete eigenvalues only, {\it i.e.}
$$\sigma(M)\cap \{z\in\C \ | \ |z|=\mbox{ \em spr}\,(M)\}\subset
\sigma_{{\rm d}}(M).$$ Then,  $$\mbox{\em spr}\,(M)\leq 1$$ and
the eigenvalues of modulus one, if any, are semisimple.
\end{prop}

\noindent
{\bf Proof:} Let $\{\alpha_j\}_{j=1,\cdots, N}$, be the discrete eigenvalues such that $|\alpha_j|=\alpha=\mbox{ spr}\,(M)$ and let  $P_j$ and $D_j$ be the corresponding eigenprojectors and eigennilpotents. Recall that $[D_j,P_j]=0$ and $P_jP_k=\delta_{jk}P_j$, for all $j,k \in \{1,\cdots, N\}$.
Setting $Q=\un -\sum_{j=1}^NP_j$, we can write, by assumption ii)
\be\label{spedec}
M=\sum_{j=1}^N\alpha_jP_j+D_j + QMQ,
\ee
where
$$
\|(QMQ)^n\|\leq e^{\beta n} \ \ \ \mbox{with}\ \ \
\beta<\ln \alpha.
$$
Let $K\in\N^*$ be such that $D_j^{K+1}=0$ for all $j\in\{1,\cdots, N\}$ and $D_{j_0}^K\neq 0$, for some $j_0\in\{1,\cdots, N\}$. If all eigennilpotents are zero, we set $K=0$. Using the properties of the spectral decomposition (\ref{spedec}), we get for any $n\in\N$ large enough
$$
M^n=\sum_{j=1}^N \left(\alpha_j^nP_j+\sum_{k=1}^K\begin{pmatrix}
n\cr k
\end{pmatrix}\alpha_j^{n-k}D_j^k\right)
 + (QMQ)^n.
$$

Consider first the case $K=0$, where all $D_j=0$. Assume that $\alpha>1$ and consider $\ffi\in {\cal H}$ such that $P_{j_0}\ffi\neq 0$, for some $j_0\in \{1,\cdots, N\}$. We define $\ffi_0=P_{j_0}\ffi/\|P_{j_0}\ffi\|$ such that $M^n\ffi_0=\alpha_{j_0}^n\ffi_0$. Now, ${\cal C}$ being dense, for any $\eps >0$, there exists $\tilde \ffi_0\in {\cal C}$ with
$\|\tilde\ffi_0-\ffi_0\|\leq \eps$ so that
$$
M^n \tilde \ffi_0=M^n\ffi_0+\sum_{j=1}^N \alpha_j^nP_j(\tilde \ffi_0-\ffi_0)+(QMQ)^n(\tilde \ffi_0-\ffi_0),
$$
where the norm of the last two terms is bounded by
$
\alpha^n \eps\left(\sum_{j=1}^N\|P_j\|  + e^{(\beta-\ln \alpha) n}\right).
$
Hence,
$$
\bra \tilde \ffi_0 | M^n \tilde \ffi_0\ket =\alpha_{j_0}^n (\bra \tilde \ffi_0 | \ffi_0\ket+O(\eps)), \ \ \ \mbox{with } \ \ O(\eps) \ \ \mbox{uniform in $n$},
$$
and $\bra \tilde \ffi_0 | \ffi_0\ket=1+O(\eps)$. Thus the modulus
of the RHS goes to infinity exponentially fast with $n$ (since
$|\alpha_{j_0}|=\alpha>1$), whereas the LHS should be uniformly
bounded in $n$ by assumption i).

Consider now $K>0$ and let $\ffi\in {\cal H}$ be such that
$D_{j_0}^K\ffi\neq 0$. Assume $\alpha\geq 1$ and set, as above,
$\ffi_0=P_{j_0}\ffi/\|P_{j_0}\ffi\|$. We have for $n$ large enough
$$
M^n\ffi_0=\alpha_{j_0}^n \left(\ffi_0 +\sum_{k=1}^K\begin{pmatrix}
n\cr k
\end{pmatrix}\alpha_{j_0}^{-k}D_j^k\ffi_0\right),
$$
where, for $1\leq k\leq K$ and  $n$ large,
$$
\begin{pmatrix} n\cr k \end{pmatrix}< \begin{pmatrix} n\cr K\end{pmatrix}
\simeq n^K/K! \ .
$$
Let $\psi_0=D_{j_0}^K\ffi_0/\|D_{j_0}^K\ffi_0\|^2$, and, for any $\eps>0$,
$\tilde \ffi_0$, $\tilde \psi_0$ in ${\cal C}$ such that $\|\tilde \ffi_0 -\ffi_0\|<\eps$ and $\|\tilde \psi_0 -\psi_0\|<\eps$. Then, as $n\ra \infty$,
\begin{eqnarray*}
\bra \psi_0 | M^n\ffi_0\ket &=& \alpha_{j_0}^n \left(
\begin{pmatrix} n\cr K\end{pmatrix}\alpha_{j_0}^{-K} + \sum_{k=1}^{K-1}\begin{pmatrix}
n\cr k
\end{pmatrix}\alpha_{j_0}^{-k}\bra \psi_0 | D_j^k\ffi_0\ket + \bra\psi_0|\ffi_0\ket
\right)\\
&=&\alpha_{j_0}^n \alpha_{j_0}^{-K} \begin{pmatrix} n\cr K\end{pmatrix}(1+O(1/n)).
\end{eqnarray*}
Thus
$$
\bra \tilde \psi_0 | M^n \tilde \ffi_0\ket= \bra \psi_0 | M^n\ffi_0\ket+
\bra \tilde \psi_0-\psi_0 |  M^n \tilde \ffi_0\ket+\bra \psi_0 | M^n (\tilde \ffi_0-\ffi_0)\ket,
$$
where the vector
$$
 M^n \tilde \ffi_0=\sum_{j=1}^N \left(\alpha_j^nP_j+\sum_{k=1}^K\begin{pmatrix}
n\cr k
\end{pmatrix}\alpha_j^{n-k}D_j^k\right)\tilde \ffi_0
 + (QMQ)^n\tilde \ffi_0
$$
satisfies for $n$ large enough and some constant $C$ uniform in $n$,
$$
\| M^n \tilde \ffi_0\|\leq C \alpha^n \begin{pmatrix} n\cr K\end{pmatrix}
\|\tilde \ffi_0\|\leq  C \alpha^n \begin{pmatrix} n\cr K\end{pmatrix}(1+\eps),
$$
and a similar estimate holds for $\| (M^n)^* \psi_0\|$. We finally get, for some constant $\tilde C$, uniform in $n$ and $\eps$,
\begin{eqnarray*}
|\bra \tilde \psi_0 | M^n \tilde \ffi_0\ket|&\geq&
|\bra \psi_0 | M^n\ffi_0\ket|-C\eps  \alpha^n \begin{pmatrix} n\cr K\end{pmatrix}
(\|\tilde \ffi_0\|+\|\psi_0\|)\\
&=&\alpha^{n-K} \begin{pmatrix} n\cr K\end{pmatrix}(1-\tilde C(1/n+\eps)).
\end{eqnarray*}
Again, if $\alpha\geq 1$, the RHS diverges as $n\ra \infty$ whereas the LHS should be bounded by ii), and the result follows.
\ep

\medskip

{\bf Remark: } To get Lemma \ref{spr}, from this Proposition, note that
 $\Psi_{\cal SR}$ is cyclic for  ${\frak M}_\cS\otimes{\frak M}_{\cR}$ and the set of analytic observables $A_{\cal SR}$ is (strongly) dense in ${\frak M}_\cS\otimes{\frak M}_{\cR}$. Moreover, Lemma \ref{quasibdd} shows that  the dense set of analytic vectors of the form $\{A_{\cal SR}\Psi_{\cal SR}\}$ satisfies assumption i). Finally, as $\Psi_{\cal SR}$ is invariant by $M_{\theta}(\lambda)$, the spectral radius is equal to $1$.

\end{document}